\documentclass[aps,prl,twocolumn,nofootinbib,groupedaddress,superscriptaddress
,longbibliography]{revtex4-2}

\usepackage{amsmath,amssymb}
\usepackage{graphicx}
\usepackage{multirow}
\usepackage{bm}
\usepackage{mathtools}
\usepackage{dsfont}
\usepackage{amsfonts}
\usepackage{xcolor}
\usepackage[normalem]{ulem}
\usepackage{url}
\usepackage{wasysym}
\usepackage{bbm}
\usepackage{hyperref}
\usepackage{braket}
\usepackage{booktabs}

\begin{document}

\title{Flux Response of Rotation-Invariant Topological Insulators}

\author{Yechen Xun}
\address{Department of Physics and Astronomy, The University of Tennessee, Knoxville, Tennessee 37996, USA}

\author{Rui-Xing Zhang}
\email{ruixing@utk.edu}
\address{Department of Physics and Astronomy, The University of Tennessee, Knoxville, Tennessee 37996, USA}
\address{Department of Materials Science and Engineering, The University of Tennessee, Knoxville, Tennessee 37996, USA}

\begin{abstract}
     Threading magnetic flux into topological phases can induce bound states that reveal intrinsic properties of the ground state. In a 3D $\mathbb{Z}_2$ topological insulator, a quantized $\pi$ flux traps a pair of 1D helical modes, whereas a trivial insulator hosts none. In this work, we show that in the presence of even-fold rotation symmetry $C_n$, a 3D band insulator features a refined $\mathbb{Z}_2 \times \mathbb{Z}_2$ classification of the flux response. Specifically, it can host two distinct types of helical flux-bound modes that are distinguished by their angular momentum. When both types of flux modes coexist, the system is not a strong topological insulator, but a $C_n$-protected topological crystalline insulator. Building on this result, we propose that flux-threaded nanowires of such topological phase provide a natural platform for realizing 1D crystalline topological superconductors with multiple $C_n$-protected Majorana modes.
\end{abstract}

\maketitle

\textit{Introduction -} The concept of topological phases has fundamentally reshaped our understanding of quantum matter~\cite{nayak2008rmp,hasan2010colloquium,qi2011RMP,chang2023colloquium}. A key frontier is to explore how various symmetries can enable novel topological states that would otherwise be impossible~\cite{haldane1983spin,kane2005QSH,bernevig2006quantum}. Although lattice symmetries are ubiquitous in crystals, their role in stabilizing new forms of topological behavior has come to light only recently~\cite{fu2011tci,shiozaki2014topo,zhang2015topo,kruthoff2017topo,bradlyn2017topological,po2017symmetry,benalcazar2017quantized,schindler2018higher,song2020twisted}. For instance, in a three-dimensional (3D) band insulator with $n$-fold rotational symmetry $C_n$ ($n=2,4,6$), one can realize $n$ gapless Dirac cones on surfaces that preserve the symmetry, while other surfaces remain gapped~\cite{fang2019new,song2017rotation}. At the intersection of neighboring $C_n$-breaking surfaces, 1D gapless hinge modes emerge. This striking boundary phenomenology, known as the rotation anomaly, stands in sharp contrast to the more uniform surface states of strong topological insulators (TIs) such as Bi$_2$Se$_3$~\cite{zhang2009topological}. The rotation anomaly further exemplifies the broader class of topological crystalline insulators (TCIs), whose topological properties are protected by crystalline symmetries, including mirror~\cite{hsieh2012topological}, glide-mirror~\cite{liu2014topo,wang2016hourglass,shiozaki2016topo,zhang2020mobius}, inversion~\cite{khalaf2018higher}, and roto-inversion~\cite{miert2018higher}.

However, only a few TCI candidates have been experimentally confirmed to date~\cite{dziawa2012topological,tanaka2012experimental,xu2012observation,liang2017KHgSb,schindler2018bismuth,fan2021discovery}. A key challenge is that, while lattice symmetries enable novel topological phases, they also constrain the sample geometries in which the associated boundary phenomena can emerge. Specifically, if the boundary breaks the protecting symmetry, the expected gapless modes of a TCI candidate may be absent. This stringent symmetry requirement, often unmet in realistic samples, can limit the observability of hallmark boundary signatures in candidate TCIs. This raises a broader conceptual question: beyond gapless boundary modes, are there any other intrinsic signatures that can reveal crystalline topology?

In this work, we address this question for 3D $C_n$-invariant band insulators by revealing a direct correspondence between bulk crystalline topology and flux-bound states. It is known that a quantized $\pi$-flux tube in a strong TI binds a pair of gapless helical modes~\cite{rosenberg2010wormhole,ostrovsky2010interaction}. Here, we show that with $C_n$ symmetry, the wavefunctions of these flux modes encode additional topological information. Specifically, the flux-bound helical modes fall into two distinct classes, characterized by $J_z = -1/2$ and $J_z = (n-1)/2$, leading to a refined $\mathbb{Z}_2 \otimes \mathbb{Z}_2$ classification. Here, $J_z$ denotes the $z$-component angular momentum. This reveals a new possibility: a band insulator can exhibit two pairs of coexisting yet decoupled helical flux modes with different $J_z$ quantum numbers. When this occurs, the system is no longer a strong TI, but a TCI with a rotation anomaly. We derive a general flux-response theory for all $C_n$ groups and propose this angular-momentum-resolved flux signature as a bulk probe of crystalline topology. Finally, we show that introducing $s$-wave Cooper pairing into this flux-bound system provides a new recipe for realizing 1D crystalline topological superconductors with symmetry-protected doublet Majorana modes.

\begin{figure}[t]
    \includegraphics[width=0.43\textwidth]{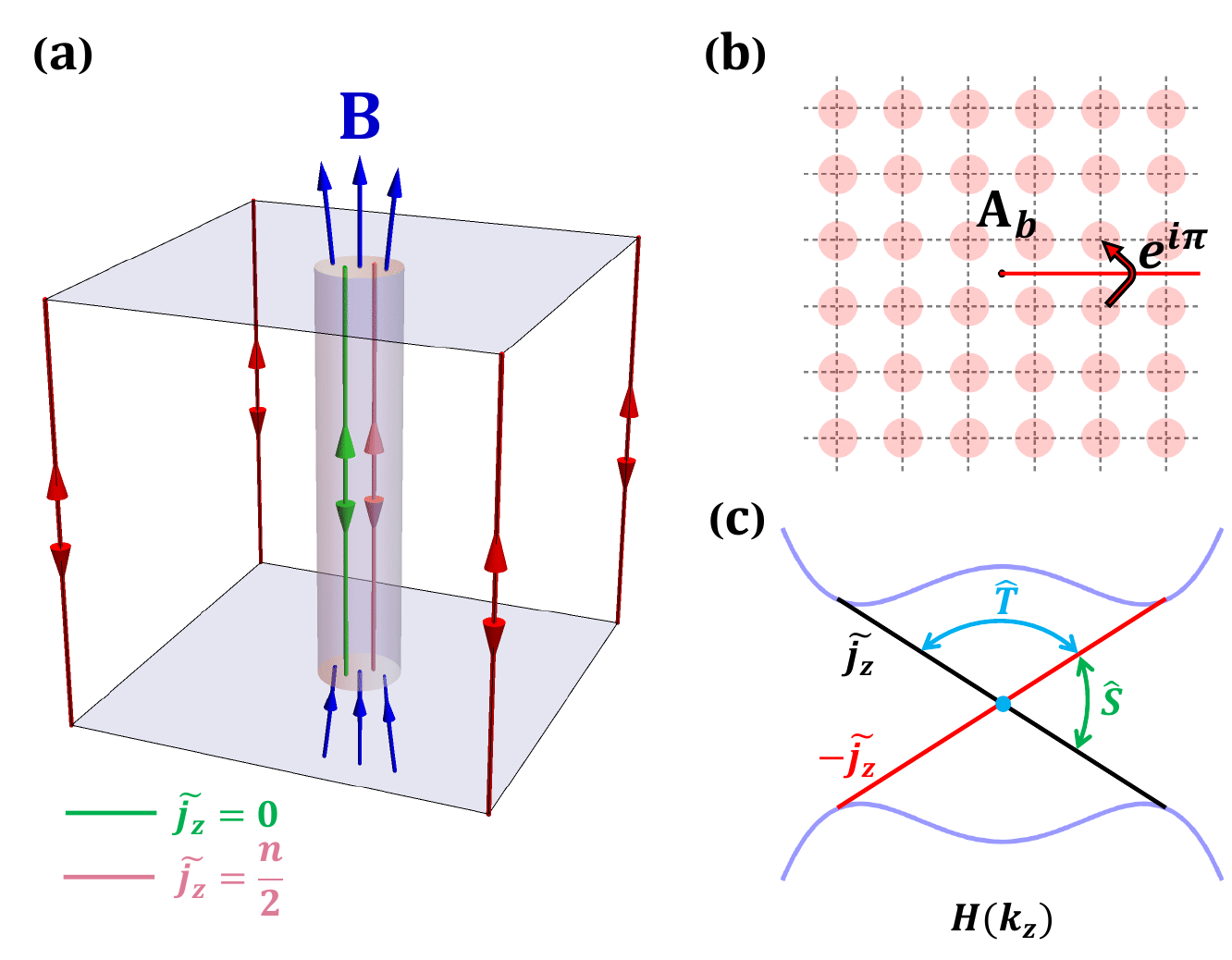}
    \caption{(a) Schematic of flux response in a $C_4$-invariant TCI. A $\pi$ magnetic flux threading the sample will trap two pairs of helical modes with $\widetilde{j_z}=0$ (green) and $n/2$ (pink). The red arrows show the helical hinge modes at the boundary. (b) Schematic of the branch-cut gauge $\bf A_b$. An extra phase $e^{i\pi}$ is obtained whenever an electron hops across the ``branch-cut" ray (red). (c) In our minimal models, the dispersion and quantum numbers of helical flux modes are constrained by $C_n$, chiral symmetry $S$, and time-reversal symmetry $T$.}
    \label{Fig1}
\end{figure}

\textit{Rotation symmetry and band topology -} We start by understanding the role of $C_n$ in topologically classifying 3D insulators from a minimalist perspective. A minimal model of 3D $\mathbb{Z}_2$ TI necessarily involves two pairs of inverted, Kramers-degenerate bands. Under $C_n$, we can exploit $J_z$ to label these bands. To preserve time-reversal symmetry (TRS) and maintain a full energy gap, we construct the basis for our model as $\Psi_{l_z} = (|l_z\rangle, |-l_z\rangle, |l_z\rangle, |-l_z\rangle)^T$, where the half-integer $l_z$ denotes the eigenvalue of $J_z$. The low-energy physics of a TI often resembles that of a massive Dirac fermion~\cite{liu2010model}. This Dirac nature requires $|l_z\rangle$ and $|-l_z\rangle$ to be coupled by $k_\pm = k_x \pm i k_y$. This leads to two inequivalent Hamiltonians for an even $n\in\{2,4,6\}$, denoted by 
\begin{align*}
    H_{\pm}({\bf k}) = Ak_z \Gamma_3 + B(k_x\Gamma_1 \pm k_y \Gamma_2) + M({\bf k}) \Gamma_5,
\end{align*}
where the Dirac $\Gamma$ matrices are defined as $\Gamma_{i=1,2,3} = s_x \otimes \sigma_i ,  \Gamma_4 = s_y \otimes \sigma_0 $, $\Gamma_5 = s_z \otimes \sigma_0 $ and $M({\bf k}) = m_0 + m_2(k_x^2+k_y^2)+m_1 k_z^2$. Both $H_+$ and $H_-$ preserve TRS $T = i s_0 \otimes \sigma_y K$ and an around-$z$ rotation $C_n = \exp[-i \frac{2\pi}{n} l_z (s_0 \otimes \sigma_z)]$. Notably, $H_+$ requires $l_z = 1/2$, while $H_-$ requires $l_z = (n-1)/2$ and is only compatible with an even-fold rotation such as $C_{2,4,6}$. The occupied states of $H_+$ and $H_-$ therefore transform differently under $C_n$ at high-symmetry points in momentum space. As a result, while both describe a $\mathbb{Z}_2$ topological insulator, $H_\pm$ are topologically distinct in the presence of $C_n$ symmetry and cannot be adiabatically connected without breaking $C_n$. The existence of two inequivalent classes of $\mathbb{Z}_2$ TIs immediately implies an enriched classification: in symmetry class AII with $C_{2,4,6}$, 3D insulators are characterized by a $\mathbb{Z}_2 \otimes \mathbb{Z}_2$ index. When both $\mathbb{Z}_2$ indices are nontrivial, the system achieves a 3D class-AII TCI with $C_n$ rotation anomaly.  

\textit{Quantum numbers of flux modes.-} We now discuss the flux response of $H_\pm$. As shown in Fig.~\ref{Fig1} (a), we consider a $\delta$-function-like magnetic flux tube ${\bf B}=\phi_0\delta(r)$ that aligns with the rotation symmetry axis of the TI, where $(r,\theta)$ denote the in-plane polar coordinates and $\phi_0 = \frac{hc}{e}$ is the flux quantum. This idealized flux profile facilitates analytical treatment, while our conclusions remain robust under more realistic flux geometries. To incorporate the effect of the flux, we use the Peierls substitution to modify the hopping term as $c^\dagger_{{\bf r}_1}c_{{\bf r}_2} \rightarrow \text{exp}(-\frac{2\pi i}{\phi_0}\int_{{\bf r}_2}^{{\bf r}_1} d {\bf r} \cdot {\bf A}) c^\dagger_{{\bf r}_1}c_{{\bf r}_2}$. There are two convenient and physically equivalent gauge choices for ${\bf A}$: the symmetric gauge ${\bf A}_s = (-y, x)/(2r^2)$ and the ``branch-cut" gauge ${\bf A}_b$ shown in Fig.~\ref{Fig1} (b). In the ${\bf A}_b$ gauge, electrons acquire a $\pi$ phase when hopping across the cut at $\theta = 0$. Both gauges satisfy the flux quantization condition $\oint_{\cal L} {\bf A} \cdot d{\bf r} = \pi$ for any closed loop ${\cal L}$ encircling the origin. As discussed in the Supplemental Material (SM)~\cite{supp}, while both gauge choices respect $C_n$, they differ in how $C_n$ is represented. ${\bf A}_s$ is compatible with $(C_n)^n = -1$, implying that the angular momentum eigenvalues $J_z = j_z$ are half-integer. In contrast, ${\bf A}_b$ yields $(C_n)^n = +1$, and $J_z = \widetilde{j_z}$ becomes integer-valued. In the following, we will use the branch-cut gauge ${\bf A}_b$ and the symmetric gauge ${\bf A}_s$ interchangeably for our purpose. To avoid ambiguity, we will distinguish between the corresponding $J_z$ eigenvalues by using different labels: $j_z$ for the symmetric gauge and $\widetilde{j_z}$ for the branch-cut gauge. As proved in the SM~\cite{supp}, they are related via 
\begin{equation} 
j_z \equiv \widetilde{j_z} - \frac{1}{2}. 
\label{eq:gauge-transform_jz}
\end{equation}

\begin{table}[t]
\centering
\renewcommand\arraystretch{1.5} 
\setlength{\tabcolsep}{1.8mm}{
\begin{tabular}{c c c c c c}
\hline \hline
   & $C_1$ & $C_2$ & $C_3$ & $C_4$ & $C_6$\\
   \hline 
   \text{Bulk Class}  & $\mathbb{Z}_2$ & $\mathbb{Z}_2 \otimes \mathbb{Z}_2$ & $\mathbb{Z}_2 $ & $\mathbb{Z}_2 \otimes \mathbb{Z}_2$ & $\mathbb{Z}_2 \otimes \mathbb{Z}_2$ \\
   $l_z$ & $N/A$ & $\frac{1}{2},\frac{1}{2}$ & $\frac{1}{2}$ & $\frac{1}{2},\frac{3}{2}$& $\frac{1}{2},\frac{5}{2}$ \\
   \hline
   \text{Flux Class}  & $\mathbb{Z}_2$ & $\mathbb{Z}_2 \otimes \mathbb{Z}_2$  & $\mathbb{Z}_2 $ & $\mathbb{Z}_2 \otimes \mathbb{Z}_2$ & $\mathbb{Z}_2 \otimes \mathbb{Z}_2$ \\
   $j_z$ & $N/A$ & $-\frac{1}{2},\frac{1}{2}$ & $-\frac{1}{2}$ & $-\frac{1}{2},\frac{3}{2}$ & $-\frac{1}{2},\frac{5}{2}$ \\
   \hline \hline
\end{tabular} }
\caption{Classification table for both bulk-state topology and helical $\pi$-flux modes with $C_n$ rotation symmetry. Specifically, the helical flux states carry an angular momentum label $j_z=-\frac{1}{2}$ only when the bulk-state basis $l_z=\frac{1}{2}$. Otherwise, the flux modes feature $j_z=l_z=\frac{n-1}{2}$ when $n$ is even. }
\label{topological classification}
\end{table}

Notably, the branch-cut gauge ${\bf A}_b$ additionally respects TRS with $T^2 = -1$, meaning the flux modes are Kramers degenerate at TRIMs. Consider a right-moving 1D flux mode with angular momentum $J_z = \widetilde{j_z}$. TRS forces its Kramers partner, a left-moving mode, to carry the opposite angular momentum, $J_z = -\widetilde{j_z}$. Additionally, both $H_\pm$ exhibit an emergent chiral symmetry ${\cal S} = \Gamma_4$, which commutes with $C_n$ but anticommutes with the Hamiltonian. As shown schematically in Fig.~\ref{Fig1} (c), this chiral symmetry forces the Kramers pairs to carry the same $J_z$ index. This leads to the condition:
\begin{equation}
    \widetilde{j_z} \equiv -\widetilde{j_z}\ (\text{mod }n)\ \Rightarrow \ \  \widetilde{j_z}=0, \frac{n}{2}.
    \label{eq:flux-jz}
\end{equation}
Namely, when $n$ is even, a pair of helical flux modes must carry a $C_n$ label with $\widetilde{j_z} = 0$ or $\widetilde{j_z} = n/2$, corresponding to a $\mathbb{Z}_2 \otimes \mathbb{Z}_2$ classification of the flux modes. However, for a $C_3$-symmetric TI, only the $\widetilde{j_z} = 0$ solution is compatible, leading to a $\mathbb{Z}_2$ classification. 

We have derived Eq.~\ref{eq:flux-jz} by invoking $T$ and ${\cal S}$. While these emergent symmetries may be absent in more realistic TI models, their absence should {\it not} affect the number or quantum numbers of the flux modes. We therefore expect Eq.~\ref{eq:flux-jz} to hold more generally. Moreover, switching from the branch-cut gauge to the symmetric gauge simply shifts the angular momentum labels from $\widetilde{j_z}$ to $j_z$, as related by Eq.~\ref{eq:gauge-transform_jz}, without affecting the underlying physics. Notably, the $\mathbb{Z}_2 \otimes \mathbb{Z}_2$ classification of the flux modes exactly reproduces the bulk $\mathbb{Z}_2 \otimes \mathbb{Z}_2$ topological classification~\cite{fang2019new}, implying a one-to-one correspondence between the two. This establishes our central result: the flux response serves as a definitive probe of crystalline topology in 3D $C_n$-invariant class-AII insulators.

\textit{Analytical and numerical solutions of flux modes.-} A direct prediction of the above ``bulk-flux correspondence" is that the flux-bound states of $H_+$ and $H_-$ should carry distinct $J_z$ indices. We now prove this result by analytically solving for the flux-bound states of $H_\pm$, respectively. We place $H_+$ and $H_-$ on a cylindrical geometry with radius $R$, where the in-plane momenta are given by $k_\pm = e^{\pm i\theta}(k_r \pm i k_\theta)$ with $k_r = -i\partial_r$ and $k_\theta = -i\partial_\theta/r$. Since the flux-bound states can be viewed as “inner boundary modes”, we instead analyze the outer boundary modes at $r = R$, which are physically equivalent but mathematically more tractable. The surface states from $H_\pm$ are labeled as $\psi_{\alpha \pm}(m)$, respectively, where $m \in \mathbb{Z}$ denotes the orbital quantum number contributing to the total angular momentum $J_z$ and $\alpha=1,2$ labels the Kramers pair. As shown in the SM~\cite{supp}, the energy dispersion of these surface states at $k_z = 0$ is
\begin{align}
    & E_{1+} = \frac{B}{R} (j_{1+}+l), \quad E_{2+} = -\frac{B}{R} (j_{2+}+l), \nonumber \\
    & E_{1-} = -\frac{B}{R} (j_{1-} - \frac{n}{2}+l), \quad E_{2-} = \frac{B}{R} (j_{2-} - \frac{n}{2}+l).
    \label{eq:surface_E}
\end{align}
Under the symmetric gauge ${\bf A}_s$, $J_z$ indices for the surface states are
\begin{align}
    & j_{1+} = m + \frac{1}{2} , \quad j_{2+} = -(m + \frac{1}{2}), \nonumber \\
    & j_{1-} = (m -\frac{1}{2}) + \frac{n}{2} , \quad j_{2-} = -(m -\frac{1}{2})+\frac{n}{2}.
    \label{eq:surface_jz}
\end{align}
In the flux-free limit ($l=0$), all surface modes exhibit a universal finite-size gap of $B/R$~\cite{rosenberg2010wormhole}.

Inserting a $\pi$-flux ($l = \frac{1}{2}$) enables zero-energy solutions in Eq.~\ref{eq:surface_E}. Including their $k_z$ dispersions, the zero modes extend to a single pair of 1D helical states along the flux line with the {\it same} $J_z$ label. As expected, the flux modes from $H_+$ and $H_-$ carry distinct $J_z$ indices,
\begin{equation} 
j_z^{(+)} = -\frac{1}{2}, \quad j_z^{(-)} = \frac{n-1}{2},
\label{eq:flux-jz-analytical}
\end{equation}
where $j_z^{(\pm)}$ denotes the $J_z$ of helical flux modes from $H_\pm$. The above relation is in exact agreement with the flux response prediction in Eqs.~\ref{eq:gauge-transform_jz} and~\ref{eq:flux-jz}. 

We can further consider a composite system, 
$H(\vec{k}) = H_+(\vec{k}) \oplus H_-(\vec{k}) + H_c(\vec{k})$,
where $H_c(\vec{k})$ introduces symmetry-preserving hybridizations between the two TI sectors. As shown in the SM~\cite{supp}, this construction realizes a $C_n$-invariant TCI with rotation anomaly. We naturally expect the TCI phase to exhibit two coexisting and $C_n$-protected pairs of helical modes bound to a $\pi$-flux tube, following both the bulk and flux classifications. 

As a numerical check, we regularize the composite Hamiltonian $H({\bf k})$ on a $C_4$-invariant cubic lattice with $k_i \rightarrow \sin k_i$ and $k_i^2 \rightarrow 2(1-\cos k_i)$ for $i\in\{x,y,z\}$. We impose open boundary conditions along both the $x$ and $y$ directions on a $20 \times 20$ lattice, while keeping $k_z$ as a good quantum number. As detailed in the SM~\cite{supp}, the symmetry-compatible hybridization term is found to be $H_c({\bf k}) = g(k_x^2 - k_y^2)\tau_y \otimes s_x \otimes \sigma_y$. We further introduce a potential shift term $\delta \tau_z \otimes s_0 \otimes \sigma_0$ that offsets the onsite energies of $H_\pm$ blocks. We choose the following model parameters: $m_1 = m_2 = -m_0 = B = 1$, $A = g = 0.5$, and $\delta = 0.1$, under which both $H_+$ and $H_-$ independently realize a strong TI phase.

\begin{figure}[t]
    \includegraphics[width=0.48\textwidth]{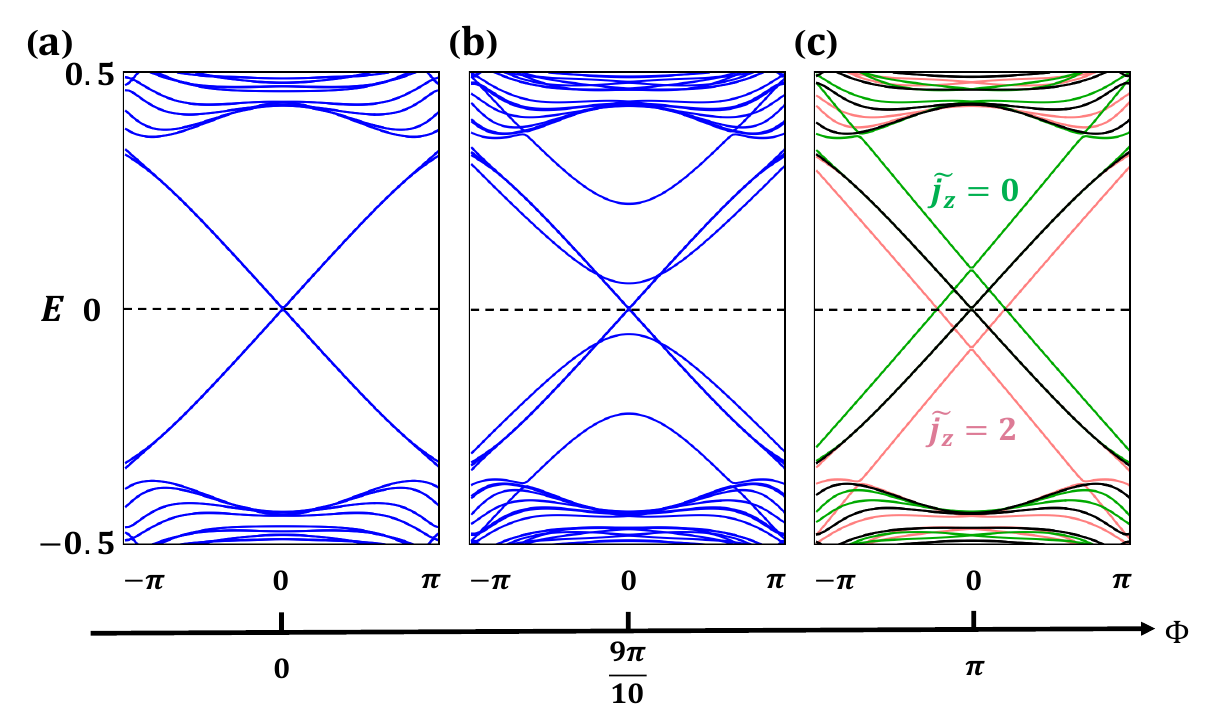}
    \caption{ The energy spectra of $H(k_z)$ on a $C_4$-symmetric lattice with open boundary conditions, as a function of flux $\Phi$. (a) shows a surface gap and gapless hinge modes. (b) with an increased flux $\Phi$ pulls flux modes into the surface gap. (c) shows two pairs of decoupled helical flux modes with $\widetilde{j_z}=0$ (green) and $n/2$ (pink), respectively.}
    \label{Fig2}
\end{figure}

Fig.~\ref{Fig2} presents the $k_z$-resolved energy spectra of $H({\bf k})$ under various magnetic flux values $\Phi$, implemented via Peierls substitution using the branch-cut gauge ${\bf A}_b$. At zero flux ($\Phi = 0$), Fig.~\ref{Fig2} (a) shows an overall gapped surface spectrum with four pairs of gapless states residing in the bulk gap. After visualizing their spatial profiles, we confirm them as 1D helical modes trapped around the ``hinges" between (100) and (010) surfaces, a characteristic of the $C_4$-symmetric TCI phase. Upon gradually increasing the flux, the surface gap remains open, while four {\it additional flux-bound states} are pushed into the surface gap, as illustrated in Fig.~\ref{Fig2} (b). When the flux reaches $\Phi = \pi$, these modes evolve into gapless helical states localized around the flux core. In Fig.~\ref{Fig2} (c), we further extract the $J_z$ quantum numbers of these flux modes and find them to be $\widetilde{j_z}=0,2$, respectively. This numerical result directly confirms our analytical predictions given in Eqs.~\ref{eq:flux-jz} and~\ref{eq:flux-jz-analytical}.    

\textit{$C_n$-enriched Majorana physics.-} An immediate and profound application of flux-response physics is the realization of non-Abelian Majorana zero modes (MZMs)~\cite{kitaev2001unpaired,read2000paired}. In 1D, a class-D topological superconductor (TSC) can be engineered from conventional $s$-wave pairing only if the normal state is effectively spinless, typically achieved by lifting spin degeneracy at the Fermi level~\cite{lutchyn2010majorana,oreg2010helical}. In a TI nanowire threaded by magnetic flux, this requirement is naturally met when a single pair of helical flux modes emerges, enabling one MZM at each wire end upon superconducting proximity~\cite{cook2011majorana}. A $C_n$-invariant TCI nanowire, however, generally supports two pairs of flux modes. This doubling of low-energy channels suggests the possibility of realizing two decoupled MZMs at each wire end when proximitized, each protected by its distinct $j_z$. If achieved, such a state would constitute a novel {\it crystalline TSC} protected by $C_4$ symmetry.

To confirm this expectation, we now turn to a numerical simulation of the proximitized TCI nanowire. We place the $C_4$-invariant TCI Hamiltonian $H({\bf k})$ on a $L \times L$ square lattice in the $x$-$y$ plane with $L = 10$, while treating $k_z$ as a good quantum number. To better capture experimental configuration, we apply a uniform magnetic field ${\bf B} = \nabla \times {\bf A}$ using the symmetric gauge ${\bf A} = l\Phi_0(-y, x)/(2L^2)$ via Peierls substitution. In the absence of flux, the four pairs of helical hinge modes identified in Fig.~\ref{Fig2} (a) are hybridized and gapped for the $L$ we choose~\cite{footnote1}. However, when a $\pi$-flux is applied, two pairs of gapless helical modes reappear. Unlike in the idealized branch-cut gauge used in Fig.~\ref{Fig2} (c), where the modes are tightly localized around the flux core, here the flux modes are redistributed along the outer boundary of the wire. Discussions on the localization length of hinge modes are provided in the SM~\cite{supp}.    

\begin{figure}[tp]
    \includegraphics[width=0.48\textwidth]{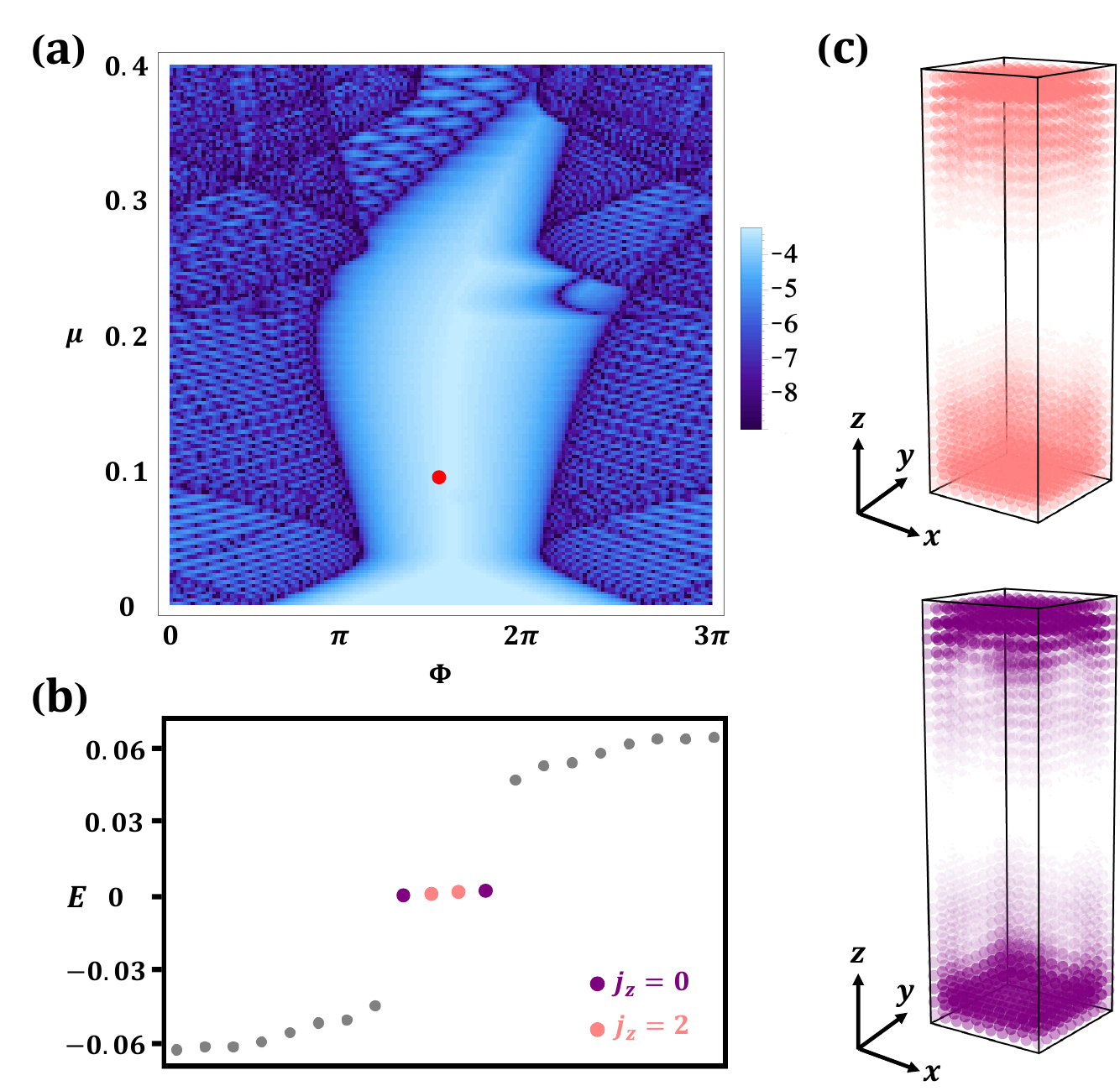}
    \caption{(a) Logarithmic color map of BdG gap for a $10 \times 10$ $C_4$-invariant TCI nanowire as a function of magnetic flux $\Phi$ and chemical potential $\mu$, with parameters $\delta = g = 0.1$ and $\Delta_0 = 0.02$. A fully gapped superconducting phase appears near $\Phi = 3\pi/2$. (b) BdG energy spectrum of a finite $10 \times 10 \times 30$ nanowire with $\mu=0.1$ and $\Phi=3\pi/2$ [red dot in (a)], showing four zero-energy Majorana modes. (c) Local density of states of MZMs in (b). MZMs with $j_z = 0$ and $j_z = 2$ are shown in purple and pink, respectively.
} 
    \label{Fig3}
\end{figure}

We begin by updating the nanowire Hamiltonian $H_L(k_z)$ to the Bogoliubov-de Gennes (BdG) form,
\begin{align*}
   {\cal H}(k_z) = \begin{pmatrix}
        H_L(k_z)-\mu && \Delta \\
        \Delta^\dagger && -H^*_L(-k_z)+\mu
    \end{pmatrix},
\end{align*}
where $\mu$ is the chemical potential. The conventional $s$-wave spin-singlet pairing is described by $\Delta = i\Delta_0 \sigma_y$, which fully gaps out all low-energy electronic states and results in a trivial BdG ground state when $l=0$. As we gradually introduce a magnetic field, however, the system begins to exhibit gapless superconducting phases, marked by zero-energy crossings of BdG bands with distinct $j_z$ labels. As discussed in Ref.~\cite{de2019conditions}, a fully gapped TSC phase can be realized only when the pairing order acquires a vortex phase winding, i.e., $\Delta(r,\theta) = i\sigma_y \Delta_0 e^{-i\theta}$, where the chirality of the vortex aligns with $\mathbf{B}$. This naturally emerges in full-shell nanowires with a symmetric superconducting coating. Crucially, this vortex aligns the $j_z$ of the crossing modes, enabling them to hybridize and open a topological gap~\cite{supp}.

By incorporating the vortex profile, we track the evolution of the minimal BdG gap across all $k_z$ values on a $10\times10$ lattice as a function of $\mu$ and $\Phi$. This yields the topological phase diagram shown in Fig.~\ref{Fig3} (a), which separates into two distinct regions: (i) a fully gapped phase (light blue) centered around $\Phi \in (\pi, 2\pi)$ and (ii) gapless regions (dark blue) characterized by multiple zero-energy band crossings. Interestingly, the maximum gap appears near $\Phi = 3\pi/2$, rather than at $\pi$ as predicted by continuum models. This shift arises from finite-size effects due to the small cross section ($L = 10$) of the nanowire.

To probe the nature of the gapped phase, we impose open boundary conditions along $\hat{z}$ with $L_z = 30$, and compute the BdG spectrum for a finite-length nanowire under a $3\pi/2$ flux [red dot in Fig.~\ref{Fig3} (a)], using $\mu = 0.1$ and $\Delta_0 = 0.05$. As shown in Fig.~\ref{Fig3} (b), four zero-energy modes appear. The spatial profiles of these modes, plotted in Fig.~\ref{Fig3} (c), confirm that all are localized at the wire ends. We numerically find that two modes carry $j_z = 0$, while the other two have $j_z = 2$. This spectrum is consistent with the $\mathbb{Z}_2 \times \mathbb{Z}_2$ classification of 1D class-D systems with $C_4$ symmetry~\cite{kobayashi2020double,hu2023topological,pandey2025QD}, which allows for up to two symmetry-distinct MZMs per end. Our setup thus realizes a minimal platform for a $C_4$-protected crystalline TSC. More broadly, nanowires made from $C_n$-invariant TCIs should enable a general route to 1D crystalline TSCs hosting a pair of decoupled MZMs with $j_z = 0$ and $j_z = n/2$, when $n$ is even.

{\it Conclusions. -} We have demonstrated, through both analytical and numerical studies, that a quantized $\pi$ magnetic flux provides an unambiguous probe of the bulk topology in $C_n$-invariant 3D band insulators. Beyond simply identifying the dispersion of flux-bound states, we show that their angular momentum quantum numbers encode a refined $\mathbb{Z}_2 \times \mathbb{Z}_2$ structure. Building on this insight, we further show that the flux response naturally guides the realization of crystalline TSC phases in nanowires proximitized by conventional $s$-wave pairing.

Meanwhile, $C_2$-protected TCI phases have been predicted in a broad range of materials, including Ca$_2$As~\cite{zhou2018C2}, Bi~\cite{hsu2019topology}, Ba$_3$Cd$_2$As$_4$\cite{zhang2019C2}, BiSe\cite{cao2021first}, and SrPb~\cite{fan2021discovery}. Nanowires of these candidates are expected to exhibit Aharonov-Bohm oscillations in magnetotransport from the periodic appearance and disappearance of gapless flux bound states. This effect should resemble that observed in Bi$_2$Se$_3$ nanowires~\cite{peng2010aharonov,bardarson2010AB}. Particularly promising is Sr$_3$SnO, which combines intrinsic superconductivity with $T_c = 4.9$~K~\cite{oudah2016superconductivity} and $C_4$-protected crystalline topology~\cite{hsieh2014SSO,kawakami2018PRX,fang2020higher,chen2025intertwining}. A superconducting Sr$_3$SnO nanowire thus offers a natural and versatile platform to experimentally achieve our prediction of flux-induced Majorana modes protected by $C_4$ symmetry.

Finally, we note that flux insertion has also been employed to probe higher-order topological phases in other symmetry classes~\cite{teo2013exist,benalcazar2014class,li2020fractional,geier2021bulk,liu2021bulk,you2023anomalous,zijderveld2024scattering}, though the resulting flux-bound states can differ markedly from those discussed here. For example, inversion-protected second-order TIs host gapped 1D flux modes and localized 0D end states at the surface termination of the flux line~\cite{schindler2022topological}, in stark contrast to the gapless flux-bound channels revealed in our work. While recent theoretical advances have established comprehensive frameworks for classifying band topology~\cite{kruthoff2017topo,bradlyn2017topological,po2017symmetry}, a general theory that seamlessly integrates crystalline symmetries, bulk topology, and flux response remains an exciting open question. We leave this intriguing direction for future exploration.

\bibliography{ref}

\begin{thebibliography}{65}
\bibitem [{foo()}]{footnote1}%
  \BibitemOpen
  \bibinfo {note} {The localization length of hinge modes $\xi$ can be
  estimated by the ratio between Fermi velocity $v_F$ and surface gap $E_g$,
  where $v_F\sim 10^5$ m/s and $E_g\sim 10^{-2}$ eV. This yields $\xi$ on the
  order of tens of nanometers.}\BibitemShut {Stop}%
\end{thebibliography}

\onecolumngrid
\appendix

\subsection{\large{Supplemental Material for ``Flux Response of Rotation-Invariant Topological Insulators"}}

\section{$C_n$ representations for ${\bf A}_b$ and ${\bf A}_s$}

\begin{figure}[hbtp!]
    \includegraphics[scale=0.4]{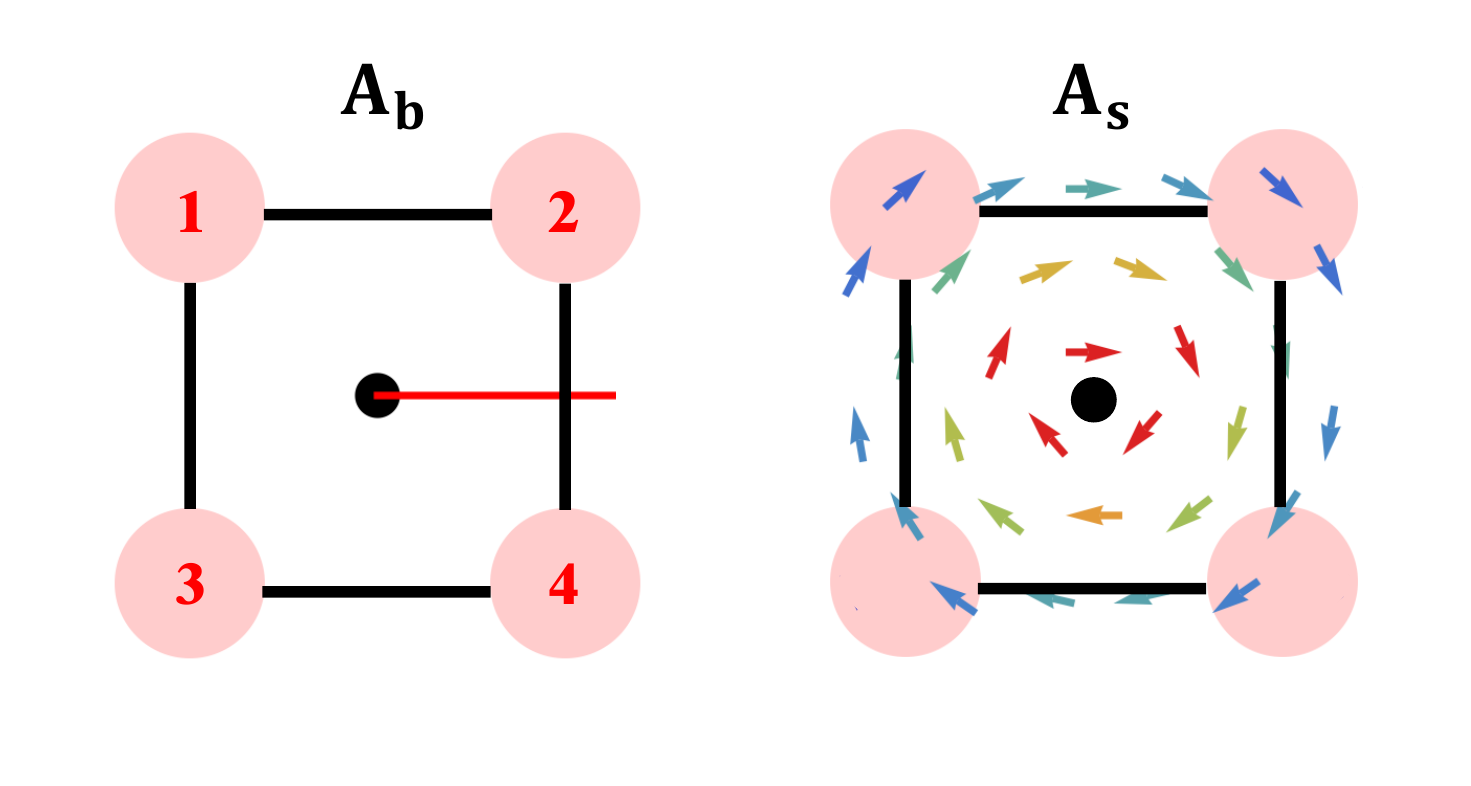}
    \caption{The branch-cut gauge ${\bf A}_b$ (left) and the symmetric gauge $\bf A_s$ (right) of the $\pi-$flux magnetic field $\bf B$ in a four-point square lattice. The red ray of ${\bf A}_b$ represents the ``branch-cut'' in 2D lattice plane through which the electrons acquire additional $\pi$ phase and the lattice points are labeled by $\{ 1,2,3,4\}$. The arrows schematically represent the vector field ${\bf A}_s = (-y,x)/(2r^2)$.}
    \label{Fig.A1}
\end{figure}

In this section, we construct the matrix representations of $C_n$ under both the branch-cut gauge ${\bf A}_b$ and the symmetric ${\bf A}_s$ for a point-like, $\pi$-quantized magnetic flux for our minimal 4-band TI model. Without loss of generality, we consider a minimal $C_4$-symmetric square lattice with four lattice points and introduce the point $\pi$-flux at its center, as shown in Fig.~\ref{Fig.A1}. For the ${\bf A}_s=\frac{1}{2}\partial_i\theta= (-y,x)/(2r^2)$, the representation matrix of $C_4$ is:
\begin{equation}
    (C_{4,s})_{ij}= \langle {\bf r}_i,\Psi_{l_z} | C_{4} | {\bf r}_j,\Psi_{l_z} \rangle = \delta({\bf r}_i - R_{4}{\bf r}_j) \langle \Psi_{l_z} | e^{-i{J_z \frac{\pi}{2}}} | \Psi_{l_z} \rangle, 
\end{equation}
Here ${\bf r}_{i,j}$ represents the position of lattice points and $\delta$ is the delta function centered at $\Delta {\bf r}=0$. $R_{n=4}$ is the operation of 4-fold rotation and $J_z$ is the angular momentum operator. The basis of this representation is $|{\bf r}_i,\Psi_{l_z} \rangle = | {\bf r}_i \rangle \otimes | \Psi_{l_z} \rangle$, where $|{\bf r}_{i=1,2,3,4} \rangle = (|1\rangle,|2\rangle,|3\rangle,|4\rangle)^T$ marks the spatial degrees of freedom and $|\Psi_{l_z} \rangle= (|l_z\rangle, |-l_z\rangle, |l_z\rangle, |-l_z\rangle)^T$ serves as the basis for the on-site Hamiltonian. The effect of rotation is reflected both in the real-space lattice transformation and the contribution of the on-site term, with $J_z$ as the generator. Then we explicitly expand the representation matrix:
\begin{equation}
    C_{4,s} = \begin{bmatrix}
        0 & 1 & 0 & 0 \\
        0 & 0 & 0 & 1 \\
        1 & 0 & 0 & 0 \\
        0 & 0 & 1 & 0
    \end{bmatrix} \otimes e^{-i\frac{\pi l_z}{2} s_0 \otimes \sigma_z},
\end{equation}
where the first $4\times4$ matrix represents the grid rotation in real space, and the exponential factor represents the on-site contribution. The total angular momentum eigenvalue associated with $C_{4,s}$ is $j_z$, whose symmetry-allowed value can be obtained through exact diagonalization of $C_{4,s}$. Since $(C_{4,s})^4 = -1$, $j_z$ must be half-integer valued, i.e., 
\begin{equation}
    j_z \in \{\pm \frac{1}{2},\pm \frac{3}{2}\}
\end{equation}

For ${\bf A}_b$ shown in Fig.~\ref{Fig.A1}, the electron acquires an additional $\pi$ phase when crossing the branch cut at $\theta=0$. As a result, $C_4$ representation compatible ${\bf A}_b$ is found to be,
\begin{equation}
    C_{4,b} \equiv \begin{bmatrix}
        0 & 1 & 0 & 0 \\
        0 & 0 & 0 & -1 \\
        1 & 0 & 0 & 0 \\
        0 & 0 & 1 & 0
    \end{bmatrix} \otimes e^{-i\frac{\pi l_z}{2} s_0 \otimes \sigma_z},
\end{equation}
where the onsite part remains the same as that of $C_{4,s}$. We denote $\widetilde{j_z}$ as the eigenvalue of the total angular momentum defined by $C_{4,b}$, to be distinguished from $j_z$ for $C_{4,s}$. It is easy to show that $(C_{4,b})^4=1$, and hence $\widetilde{j_z}$ is integer-valued:
\begin{equation}
    \widetilde{j_z}\in\{0,1,2,3\}.
\end{equation}

Since both gauge choices respect $C_4$, it is natural to ask whether and how $C_{4,s}$ and $C_{4,b}$ are related to each other. We make the following observation by exploiting the Jordan decomposition, 
\begin{equation}
    e^{-i\frac{\pi}{4}} C_{4,s} = U C_{4,b} U^{-1}, \quad U=\begin{bmatrix}
        -i_{4\times 4} & 0 & 0 & 0 \\
        0& -(-1)^{\frac{1}{4}}_{4\times 4} & 0 & 0\\
        0 & 0 & -(-1)^{\frac{3}{4}}_{4\times 4} & 0\\
        0 & 0 & 0 & 1_{4\times4} 
    \end{bmatrix}.
\end{equation}
Namely, $\widetilde{C}_{4,s}\equiv e^{-i\frac{\pi}{4}} C_{4,s}$ is {\it unitary similar} to $C_{4,b}$, and they hence should share the same eigenvalue $\text{exp}(-i\frac{\pi}{2}\widetilde{j_z})$. This naturally leads to 
\begin{equation}
   j_z =  \widetilde{j_z} - \frac{1}{2},
\end{equation}
which is Eq.~1 of the main text. The above discussion and the equivalence relation between ${\bf A}_s$ and ${\bf A}_b$ for a point-like flux quantum can be straightforwardly generalized to systems with a larger size and/or a different rotational symmetry.

\section{Surface state of $C_n$-invariant topological insulators}

In this section, we analytically derive the dispersion relations and wavefunctions for the surface states of $H_\pm({\bf k})$ under cylindrical coordinates, in the absence of any magnetic flux.

The effective $k \cdot p$ Hamiltonian $H_+({\bf k})$ of a 3D strong TI with $C_n$ and TRS is $H_{+}({\bf k}) = Ak_z \Gamma_3 + B(k_x\Gamma_1 + k_y \Gamma_2) + M({\bf k}) \Gamma_5$ defined under the basis $(|l_z\rangle , |-l_z\rangle,|l_z\rangle,|-l_z\rangle)^T$. The Dirac $\Gamma$ matrices are defined as $\Gamma_{i=1,2,3} = s_x \otimes \sigma_i ,  \Gamma_4 = s_y \otimes \sigma_0 $ and $\Gamma_5 = s_z \otimes \sigma_0 $. $l_z$ denotes the eigenvalue of $J_z$, the $\hat{z}$-component angular momentum. $H_+({\bf k})$ can be transformed into a block-diagonal form at $k_z=0$ by a unitary matrix $U$:
 \begin{equation}
	  H_{+}({\bf k}) \rightarrow U^{-1}H_{+}({\bf k})U = \begin{bmatrix}
		h_1 && 0\\
		0 && h_2 \\ 
	\end{bmatrix} ,
\end{equation}
where
\begin{equation}
    h_1 = \begin{bmatrix}
        M({\bf k}) && Bk_-\\
        Bk_+ && -M({\bf k})
    \end{bmatrix} , h_2 = \begin{bmatrix}
        -M({\bf k}) && Bk_-\\
        Bk_+ && M({\bf k})
    \end{bmatrix} , U= \begin{bmatrix}
        1 && 0 && 0 && 0\\
        0 && 0 && 0 && 1\\
        0 && 0 && 1 && 0\\
        0 && 1 && 0 && 0
    \end{bmatrix}.
\end{equation}
Here $M({\bf k})\equiv m_0+m_2(k_x^2+k_y^2)+m_1 k_z^2$ and $k_\pm = k_x \pm i k_y$, with $m_0 <0 $ and $B,m_{1,2}>0$. The low-energy eigenvalue problem now can be solved in each diagonal block. We now turn to cylindrical coordinates. The $k_\pm$ term can be rewritten as $k_\pm = e^{\pm i\theta} (k_r \pm ik_\theta)$ with $k_r = -i\partial_r$ and $k_\theta = -\frac{i}{r} \partial_\theta$. Consider the nanowire radius $R$ to be much larger than the lattice constant $a$, the Hamiltonian has the form:
\begin{align}
    U^{-1}H_{+}({\bf k}) U = H_0({\bf k})  + & \begin{bmatrix}
        0 && -\frac{B}{R}e^{-i\theta}\partial_\theta && Ak_z && 0 \\
        \frac{B}{R}e^{i\theta}\partial_\theta && 0 && 0 && -Ak_z \\
        Ak_z && 0 && 0 && -\frac{B}{R}e^{-i\theta}\partial_\theta \\
        0 && -Ak_z && \frac{B}{R}e^{i\theta}\partial_\theta && 0
    \end{bmatrix}, \nonumber \\ 
    H_0({\bf k}) = \begin{bmatrix}
        h_{+} && 0 \\
        0 && h_{-}
    \end{bmatrix}  \quad &, \quad  h_{\pm}=  \begin{bmatrix}
        \pm (m_0-m_2\partial_r^2) && -iBe^{-i\theta}\partial_r  \\
        -iBe^{i\theta}\partial_r && \mp (m_0-m_2\partial_r^2) 
    \end{bmatrix}.
\end{align}
We will solve for boundary zero modes of $H_0({\bf k})$ and treat the remaining terms as perturbations. To find the zero-energy state, we consider an ansatz solution:
\begin{align}
	 \psi = e^{im\theta}\psi_r = e^{im\theta} e^{\lambda (r-R)} \begin{bmatrix}
		f \\
		g
	\end{bmatrix}\quad , \quad r<R
\end{align}
Here $m\in\mathbb{Z}$. This immediately leads to
\begin{align}
	h_\pm \psi & = \begin{bmatrix}
		\pm (m_0-m_2\lambda^2) && -iBe^{-i\theta}\lambda\\
		-iBe^{i\theta}\lambda && \mp (m_0-m_2\lambda^2) \\
	\end{bmatrix} \begin{bmatrix}
		f\\
		g
	\end{bmatrix} =0,
\end{align}
with which we find the two solutions $\lambda_{1,2}$ with Re$(\lambda)>0$ ($B=m_2=-m_0=1$):
\begin{align}
    \lambda_{1,2} = \frac{1}{2} \pm i\frac{\sqrt{3}}{2}
\end{align}
The wavefunction now can be written as a linear superposition of $\lambda_{1,2}$ eigenstates:
\begin{align}
    \psi = C_1\psi_{\lambda_1}+C_2\psi_{\lambda_2} \sim  e^{im\theta} \text{sin}(\frac{\sqrt{3}}{2}\widetilde{r}) e^{\frac{\widetilde{r}}{2}} \begin{bmatrix}
		f \\
		g
	\end{bmatrix} \quad , \quad \widetilde{r} \equiv r-R.
\end{align}
Substituting the value of $\lambda$, we immediately obtain
\begin{equation}
    g = \pm ie^{i\theta}f .
\end{equation}
Then the two eigenfunctions of the boundary state with zero energy ($r = R$) under the original basis $(|l_z\rangle , |-l_z\rangle,|l_z\rangle,|-l_z\rangle)^T$ are:
\begin{align}
	\psi_{1+} = \frac{e^{im_1\theta}}{\sqrt{2}} \begin{bmatrix}
		1 \\ 0 \\0 \\ ie^{i\theta}
	\end{bmatrix} \quad , \quad \psi_{2+} = \frac{e^{-i(m_2+1)\theta}}{\sqrt{2}} \begin{bmatrix}
		0  \\ -e^{i\theta} \\  -i \\ 0
	\end{bmatrix},
    \label{eq:eigenfunction of H+}
\end{align}
where we have ignored the radial part for simplicity. 

Since $l_z = 1/2$ in case of $H_{+}({\bf k})$, $\psi_1$ carries total angular momentum $j_{1+} = m_1 + \frac{1}{2}$ and $\psi_2$ with $j_{2+}=-(m_2 + \frac{1}{2})$ that can be related to $\psi_1$ by time-reversal operation $is_0\otimes \sigma_y K$. Then we complete the perturbation under the basis of $|\Psi \rangle = (|\psi_1\rangle,|\psi_2\rangle )^{T}$ by projection:
\begin{align}
	H_{+,\text{eff}} = \langle \Psi | \begin{bmatrix}
		0 && 0 && Ak_z && -\frac{B}{R}e^{-i\theta}\partial_\theta\\
		0 && 0 && \frac{B}{R}e^{i\theta}\partial_\theta && -Ak_z \\
		Ak_z && -\frac{B}{R}e^{-i\theta}\partial_\theta && 0 && 0 \\
		\frac{B}{R}e^{i\theta}\partial_\theta && -Ak_z && 0 && 0
	\end{bmatrix} | \Psi \rangle =\begin{bmatrix}
	    \frac{B}{R}(m+\frac{1}{2}) && iAk_z \\
            -iAk_z && -\frac{B}{R}(m+\frac{1}{2})
	\end{bmatrix}.
\end{align}
The energy dispersion is $E=\pm \sqrt{(Ak_z)^2+(\frac{B}{R})^2(m+1/2)^2}$ where $m\equiv m_1 = -m_2-1$. The energy spectrum at $\Gamma_z$ ($k_z=0$) is gapped due to an intrinsic $\pi$ Berry phase of the Dirac fermion, from where the constant $\frac{1}{2}$ comes, and all gaps between neighboring energy levels are identical, $\Delta_{gap} = \frac{B}{R}$. All bands are found to be doubly degenerate. The degenerate bands with $E = \frac{B}{R}(m+\frac{1}{2})$ must carry opposite angular momentum with $J_z = \pm (m + \frac{1}{2})$, as enforced by the time-reversal symmetry.

The Hamiltonian of the TI with opposite helicity, $H_-({\bf k})$, is obtained by replacing $k_\pm$ with $k_\mp$. The eigenfunction and energy spectrum can be obtained using the same method:
\begin{align}
	\psi_{1-} = \frac{e^{im_3\theta}}{\sqrt{2}} \begin{bmatrix}
		1 \\ 0 \\0 \\ ie^{-i\theta}
	\end{bmatrix} \quad , \quad \psi_{2-} = \frac{e^{-i(m_4-1)\theta}}{\sqrt{2}} \begin{bmatrix}
		0 \\ -e^{-i\theta} \\  -i \\ 0
	\end{bmatrix},
\end{align} 
with $E =-\frac{B}{R}(m-\frac{1}{2})$. The values of angular momentum in this case are $j_{1-}=(m_3-\frac{1}{2})+\frac{n}{2}$ and $j_{2-}=-(m_4-\frac{1}{2})+\frac{n}{2}$ because the $C_n$ requires $l_z = (n-1)/2$. For summary, we list the energy dispersion and the corresponding $j_z$ of the surface states at $k_z=0$:
\begin{align}
    H_{+}({\bf k}) : \begin{cases}
        E_{1+} = \frac{B}{R}j_{1+}  &, \quad j_{1+} = (m_1+\frac{1}{2}). \\
        E_{2+} = -\frac{B}{R}j_{2+}  &, \quad j_{2+} = -(m_2+\frac{1}{2}).
    \end{cases} \quad , \quad H_{-}({\bf k}) : \begin{cases}
        E_{1-} = -\frac{B}{R}(j_{1-}-\frac{n}{2})  &, \quad j_{1-} = (m_3-\frac{1}{2})+\frac{n}{2}. \\
        E_{2-} = \frac{B}{R}(j_{2-}-\frac{n}{2})  &, \quad j_{2-} = -(m_4-\frac{1}{2})+\frac{n}{2}.
    \end{cases}
\end{align}

\section{Helical modes under $\pi$-flux}

Let us turn on a uniform magnetic flux $\Phi = l\Phi_0$ along $\hat{z}$, where $\Phi_0=\frac{hc}{e}$ is the flux quantum, and reevaluate the above surface-state problem. Then the electron Hamiltonian under ${\bf A}_s$ is:
\begin{align}
    H = \frac{1}{2m_e} [p_r {\bf e_r} + (p_{\theta} +\frac{e}{c} A_\theta){\bf e_\theta}]^2 + V(r,\theta) = \frac{1}{2m_e} [\hbar k_r {\bf e_r} + \hbar (k_\theta+\frac{lr}{R^2}){\bf e_\theta}]^2+V(r,\theta),
\end{align}
where $A_\theta = l\Phi_0r/2\pi R^2$. We see that the vector potential couples to ${\bf k}$, resulting in a shift of the canonical momentum along $\theta$-direction: $\widetilde{k}_{\theta}=k_\theta+lr/R^2 = -i(\partial_\theta+ilr^2/R^2)/r$. In the case of the boundary state ($r=R$), the following terms scale as $1/R$, and can thus be regarded as a perturbation in the large $R$ limit:  
\begin{align}
    \begin{bmatrix}
		0 && 0 && Ak_z && -\frac{B}{R}e^{-i\theta}(\partial_\theta+il)\\
		0 && 0 && \frac{B}{R}e^{i\theta}(\partial_\theta+il) && -Ak_z \\
		Ak_z &&  -\frac{B}{R}e^{-i\theta}(\partial_\theta+il) && 0 && 0 \\
		\frac{B}{R}e^{i\theta}(\partial_\theta+il) && -Ak_z && 0 && 0
	\end{bmatrix}.
\end{align}
The zero-mode eigenfunctions are not affected by the flux insertion and takes the same forms as Eq.~\ref{eq:eigenfunction of H+}. 

Then the dispersion relation about $k_z$ is $E=\pm \sqrt{(Ak_z)^2+(\frac{B}{R})^2(m+\frac{1}{2}+l)^2}$ labeled by $m$. The energy spectrums at $k_z=0$ are shifted into $E_1 = \frac{B}{R}(m_1+\frac{1}{2}+l),E_2 = \frac{B}{R}(m_2+\frac{1}{2}-l)$ and for the $\pi$-flux case, $l=1/2$, the $\pi$ Berry phase is exactly canceled. The zero-energy states with $m_1+1= m_2 = 0$ now form a gapless helical dispersion along $k_z$ following $E(k_z) = \pm A k_z$. Under symmetric gauge ${\bf A}_s$, by our definition, the angular momentum of each state is still a half-integer: $j_1 = (m_1+\frac{1}{2})$ and $j_2 = -(m_2+\frac{1}{2})$. The energy spectrum with the additional flux now can be represented as:
\begin{align}
H_{+}({\bf k}) : \begin{cases}
        E_{1+} = \frac{B}{R}(j_{1+}+l)  &, \quad j_{1+} = (m_1+\frac{1}{2}). \\
        E_{2+} = -\frac{B}{R}(j_{2+}+l)  &, \quad j_{2+} = -(m_2+\frac{1}{2}).
    \end{cases}
\end{align}

\begin{figure}[t]
    \includegraphics[scale=0.5]{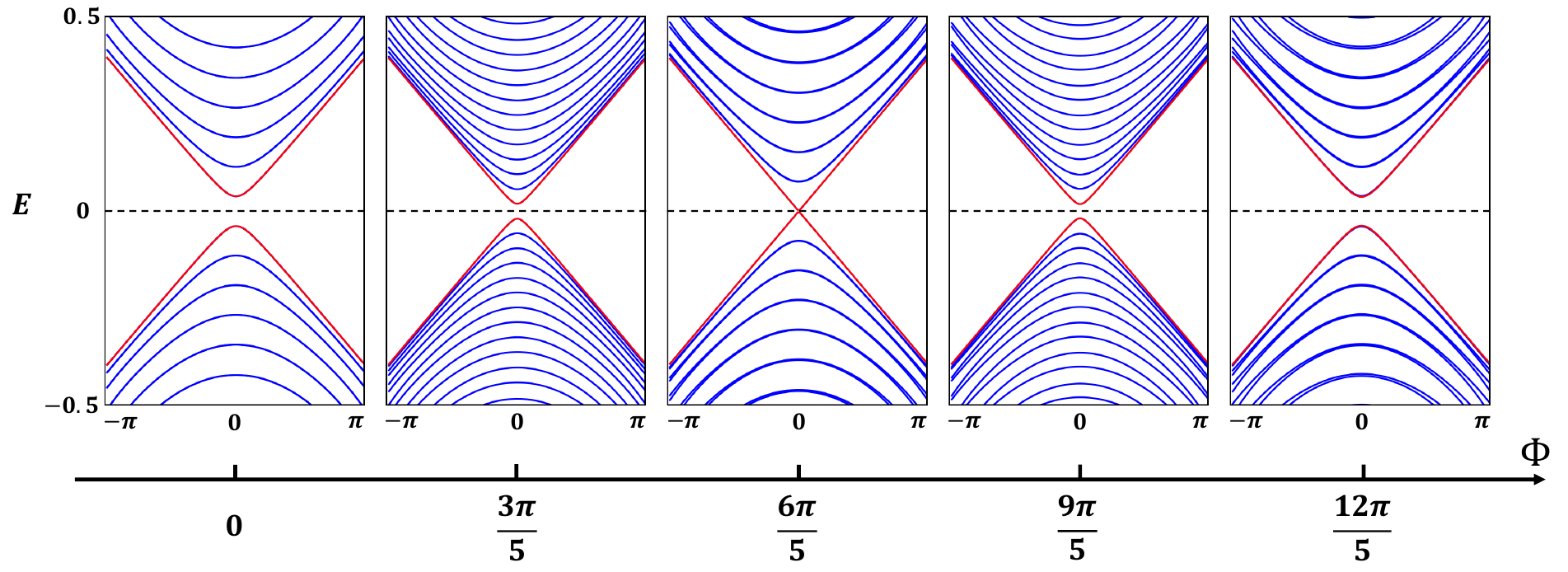}
    \caption{The energy spectrum of the 3D TI nanowire under different values of magnetic flux. The lowest energy bands are highlighted in red.}
    \label{Fig.A2}
\end{figure}

\begin{figure}[t]
     \includegraphics[scale=0.6]{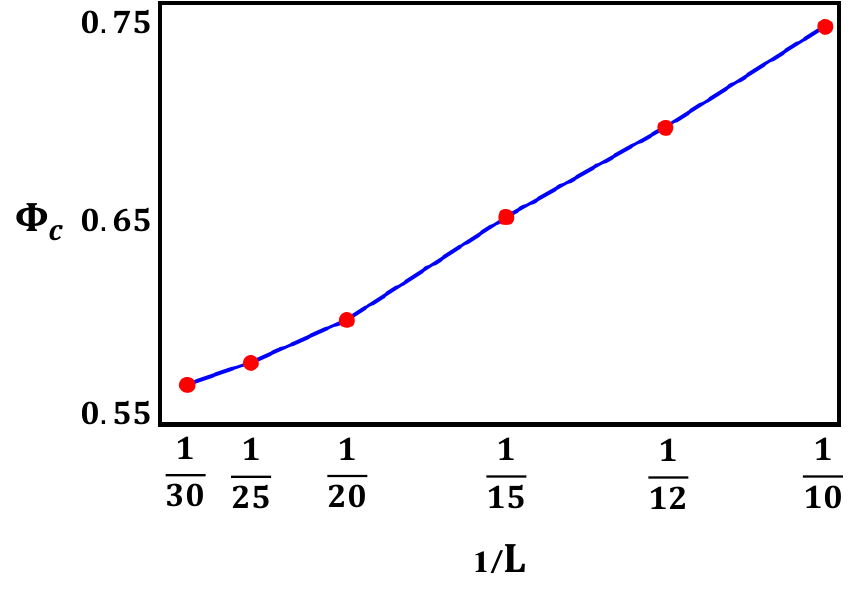}
     \caption{The critical flux $\Phi_c$ for surface gap closure as a function of $1/L$, where $L$ is the side length of the finite-size lattice.}
     \label{Fig.A3}
 \end{figure}

Switching to the branch-cut gauge $\bf A_b$ leaves the energy spectrum invariant, while updating $\widetilde{j_z} \equiv j_z +\frac{1}{2}$. We thus find that
\begin{align}
    H_{+}({\bf k},l=\frac{1}{2}) : \begin{cases} 
    E_{1+} = \frac{B}{R} \widetilde{j}_{1+} \quad &, \quad \widetilde{j}_{1+} = (m_1+1). \\
    E_{2+} = -\frac{B}{R} \widetilde{j}_{2+} \quad &, \quad \widetilde{j}_{2+} = -m_2.
    \end{cases}
\end{align}
We can clearly see that the angular momentum carried by the helical flux modes is $\widetilde{j}_{1+}=\widetilde{j}_{2+}=0$ ($j_{1+} = j_{2+} = -\frac{1}{2}$), which is consistent with our symmetry-based theoretical predictions in the main text. Since time-reversal symmetry (TRS) is respected by ${\bf A}_b$ at $\pi$-flux, this helical mode carried good quantum number $\widetilde{j_z}$ is protected by both TRS and rotational symmetry. The lattice simulation of the above discussion based on a $20 \times 20$ square lattice using the symmetric gauge ${\bf A}_s$ is shown in Fig.~\ref{Fig.A2}. The critical flux for band closure at $k_z=0$ is found to be around $6\pi/5$ due to the finite size effect. We further verify the influence of the actual lattice size on the critical flux value $\Phi_c$. Fig.~\ref{Fig.A3} shows how the critical flux (in units of $2\pi$) for gap closure depends on lattice size. As the lattice grows, we find $\Phi_c$ to approach the analytical value of $\pi$.

The flux-induced helical modes for $H_-{({\bf k})}$ can be obtained through an identical procedure:
\begin{align}
	\psi_{1-} = \frac{e^{im_3\theta}}{\sqrt{2}} \begin{bmatrix}
		1 \\ 0 \\0 \\ ie^{-i\theta}
	\end{bmatrix} \quad , \quad \psi_{2-} = \frac{e^{-i(m_4-1)\theta}}{\sqrt{2}} \begin{bmatrix}
		0 \\ -e^{-i\theta} \\ -i \\ 0
	\end{bmatrix}. \nonumber \\
    H_{-}({\bf k}):\begin{cases}
        E_{1-} = -\frac{B}{R}(j_{1-}-\frac{n}{2}+l) \quad &, \quad j_{1-} = (m_3-\frac{1}{2})+\frac{n}{2}. \\
        E_{2-} = \frac{B}{R}(j_{2-}-\frac{n}{2}+l) \quad &, \quad j_{2-} = -(m_4-\frac{1}{2})+\frac{n}{2},
    \end{cases}  
\end{align}
which gives another set of helical flux modes with $j_{1-} = j_{2-} =\frac{n-1}{2} $, equivalently $\widetilde{j}_{1-}=\widetilde{j}_{2-} = \frac{n}{2}$ under $\bf A_b$: 
\begin{align}
    H_{-}({\bf k},l=\pi):\begin{cases}
        E_{1-} = -\frac{B}{R}(\widetilde{j}_{1-}-\frac{n}{2}) \quad &, \quad \widetilde{j}_{1-} = m_3+\frac{n}{2}. \\
        E_{2-} = \frac{B}{R}(\widetilde{j}_{2-}-\frac{n}{2}) \quad &, \quad \widetilde{j}_{2-} = -(m_4-1)+\frac{n}{2}.
    \end{cases}  
\end{align}

\section{Topological crystalline insulators with $C_n$ symmetry}

In this section, we discuss how to stack the minimal TI models $H_\pm({\bf k})$ to achieve a TCI phase with rotation anomaly. We need to find the symmetry-preserving coupling term $H_c$, with which the total Hamiltonian becomes
\begin{align}
    H_{d}({\bf k}) = \begin{bmatrix}
        H_{+}({\bf k}) && 0 \\
        0 && H_{-}({\bf k})
    \end{bmatrix} + H_c({\bf k}).
\end{align}
In the following discussion, we change the basis of $H_{\pm}({\bf k})$ to $( |l_z\rangle , |l_z \rangle, |-l_z \rangle , |-l_z\rangle )^T$ for convenience:
\begin{align}
   H_{\pm}({\bf k}) \rightarrow U^{-1}H_{\pm}({\bf k}) U = M_0({\bf k}) s_0\otimes \sigma_z + Ak_z s_z\otimes \sigma_x + B(k_x s_x \otimes \sigma_x \pm k_y s_y \otimes \sigma_x) \quad , \quad U=\begin{bmatrix}
       1 & 0 & 0 & 0\\
       0 & 0 & 1 & 0\\
       0 & 1 & 0 & 0\\
       0 & 0 & 0 & 1\\
   \end{bmatrix}.
\end{align}

\subsection{$C_2$}

We start with the coupling terms between two single TIs that respect TRS ($T = i\tau_0 \otimes s_y \otimes \sigma_0 K$) and $C_2=-i\tau_0 \otimes s_z \otimes \sigma_0$ symmetry. We consider a constant coupling term, whose general form takes
\begin{align}
	H_c({\bf k}) = \tau_{x,y} \otimes s_{0,x,y,z} \otimes \sigma_{0,x,y,z}.
\end{align}
Only 8 of them are found to be symmetry compatible: $\tau_{x} \otimes s_{0} \otimes \sigma_{0,x,z}, \tau_{y} \otimes s_{0} \otimes \sigma_{y}, \tau_{y} \otimes s_{z} \otimes \sigma_{0,x,z}$ and $\tau_{x} \otimes s_{z} \otimes \sigma_{y}$.

We now explore the effective boundary physics for each surface. First, we solve the surface state of (001) based on a single TI by setting $k_x=k_y=0$ before perturbation and replacing $k_z \rightarrow -i\partial_z$. The zero-mode equation is
\begin{align}
	[m_0 s_{0} \otimes \sigma_z + m_2 k_z^2 s_{0} \otimes \sigma_z  + A k_z s_z \otimes \sigma_x]\psi = 0 \rightarrow [m_0 - m_2 \partial_z^2 +A\partial_z s_z \otimes \sigma_y] \psi = 0,
\end{align}
with an ansatz solution $\psi(z,s) = f(z)\zeta_s$ where $(s_z \otimes \sigma_y)\zeta_s = S \zeta_s, S=\pm1$, we get:
\begin{align}
	[m_0 - m_2 \partial_z^2 +AS\partial_z ] f(z)  = 0 \rightarrow f(z) = \mathcal{N} sin(\alpha x)e^{\beta x}, \label{wavefunction on surface}
\end{align}
where $\alpha = \frac{\sqrt{-4m_0m_2-(AS)^2}}{2m_2}, \beta = \frac{AS}{2m_2}$ and the normalization factor $\mathcal{N} = \sqrt{4\beta(\alpha^2+\beta^2)/\alpha^2}$. Here we impose the boundary conditions $f(0)=f(-\infty)=0$ and the form of $f(z)$ is ensured by the discriminant $(AS)^2+4m_0m_2<0$. We set $A,m_2>0$, then the decay solution along $-z$ direction requires $S=+1$. The positive sign of $S$ corresponds to the same sign conditions for the eigenvalue of $s_z$ and $\sigma_y$. Thus, the two eigenfunctions are:
\begin{align}
	\zeta_1 = \begin{bmatrix}
		1\\0
	\end{bmatrix} \otimes \begin{bmatrix}
		1\\i
	\end{bmatrix} \quad , \quad \zeta_2 = \begin{bmatrix}
		0\\1
	\end{bmatrix} \otimes \begin{bmatrix}
		1\\-i
	\end{bmatrix}.
    \label{eq:001-surf-states}
\end{align}
The lowest-order dispersion of the surface state can be found by projecting the linear-in-$k$ term $B(k_x s_x \otimes \sigma_x + k_y s_y \otimes \sigma_x) $ onto the zero-mode basis,
\begin{align}
	H_{+,(001)} = B(k_x \sigma_y -k_y \sigma_x)\quad , \quad C_2 = -i\sigma_z \quad, \quad T = i\sigma_y K.
\end{align}
The analysis above for $H_{+}({\bf{k}})$ directly applies to $H_{-}({\bf{k}})$ by simply replacing $k_y \rightarrow -k_y$. As such, the effective energy dispersion on (001) of the band insulator characterized by $H_{d}({\bf k}) = H_{+}({\bf k}) \oplus H_{-}({\bf k})$ is:
\begin{align}
    H_{-,(001)} = B(k_x \sigma_y + k_y \sigma_x)\quad &, \quad C_2 = -i\sigma_z \quad, \quad T = i\sigma_y K. \nonumber \\
    H_{d,(001)} = B(k_x \tau_0 \otimes \sigma_y -k_y\tau_z \otimes \sigma_x)\quad &, \quad C_2 = -i\tau_0 \otimes \sigma_z \quad, \quad T = i\tau_0 \otimes\sigma_y K.
\end{align}

The same procedure applies to the (100) and (010) surfaces, here we list the effective eigenfunction $\zeta$ and the linear dispersion of the surface state:
\begin{align}
	(100): \zeta_1 = \begin{bmatrix}
		1\\1
	\end{bmatrix} \otimes \begin{bmatrix}
		1\\i
	\end{bmatrix} \quad , \quad \zeta_2 = \begin{bmatrix}
		1\\-1
	\end{bmatrix} \otimes \begin{bmatrix}
		1\\-i
	\end{bmatrix} \quad , \quad H_{d,(100)} =B k_y \tau_z \otimes \sigma_x + Ak_z \tau_0 \otimes \sigma_y \nonumber \\
	(010): \zeta_1 = \begin{bmatrix}
		1\\i
	\end{bmatrix} \otimes \begin{bmatrix}
		1\\i
	\end{bmatrix} \quad , \quad \zeta_2 = \begin{bmatrix}
		1\\-i
	\end{bmatrix} \otimes \begin{bmatrix}
		1\\-i
	\end{bmatrix} \quad , \quad H_{d,(010)} =Ak_z \tau_z \otimes \sigma_y- B k_x \tau_z \otimes \sigma_x.
    \label{eq:100-surf-states}
\end{align}

To induce a full energy gap on the surface, the projected coupling term must anticommute with both kinetic terms of the surface Dirac Hamiltonian. The projections of symmetry-allowed coupling terms on each surface are shown in Table.~\ref{c2 table}. First, we see that the nonzero projections on the (001) surface are $\tau_x \otimes \sigma_0$ and $\tau_y \otimes \sigma_z$, where $\tau_x \otimes \sigma_0$($\tau_y \otimes \sigma_z$) commutes with $\tau_0 \otimes \sigma_y$($\tau_z \otimes \sigma_x$) in the $H_{d,(001)}$ and cannot lead to a full gap. Therefore,
\begin{itemize}
    \item The (001) surface remains gapless for any symmetry-preserving perturbation.
\end{itemize}

In contrast, the side surfaces (100) and (010) admit symmetry-allowed gapping terms. Specifically, $\tau_y \otimes \sigma_x$ on (100) anticommutes with both  $\tau_z \otimes \sigma_x$ and  $\tau_0 \otimes \sigma_y$, and $\tau_x \otimes \sigma_0$ on (010) anticommutes with both  $\tau_z \otimes \sigma_x$ and  $\tau_z \otimes \sigma_y$. Thus, the side surfaces can be fully gapped by introducing $\tau_y \otimes s_z \otimes \sigma_z$ and $\tau_x \otimes s_0 \otimes \sigma_z$ simultaneously, which, under the original basis $( |l_z\rangle , |-l_z \rangle, |l_z \rangle , |-l_z\rangle )^T$, correspond to $\tau_y \otimes s_z \otimes \sigma_z$ and $\tau_x \otimes s_z \otimes \sigma_0$ respectively. However, as pointed out in Ref.~\cite{fang2019new}, there must always exist pairs of 1D helical modes living on two $C_2$-related ``hinges" between $(100)$ and $(010)$ surfaces. The gapless (001) surface and the helical hinges together constitute the ``rotation anomaly" for the $C_2$-protected crystalline topology of this stacked TI system.

\begin{table}[t]
\centering
\begin{tabular}[c]{|c|ccc|c|c|}
    \hline
    Coupling Terms & (100) & (010) & (001) &\text{(100) Gap} & \text{(010) Gap}  \\
    \hline
    $\tau_x \otimes s_0 \otimes \sigma_0$ & $\tau_x \otimes \sigma_0$ & 0 & $\tau_x \otimes \sigma_0$ & $\times$ & $\times$ \\
    $\tau_x \otimes s_0 \otimes \sigma_x$ & 0 & $\tau_y \otimes \sigma_z$ & 0 & $\times$ & $\times$ \\
    $\tau_x \otimes s_0 \otimes \sigma_z$ & 0 & $\tau_x \otimes \sigma_0$ & 0 & $\times$ & $\checkmark$ \\
    $\tau_y \otimes s_0 \otimes \sigma_y$ & $\tau_y \otimes \sigma_z$ & 0 & $\tau_y \otimes \sigma_z $ & $\times$ & $\times$ \\
    $\tau_y \otimes s_z \otimes \sigma_0$ & 0 & $\tau_y \otimes \sigma_x$ & $\tau_y \otimes \sigma_z $ & $\times$ & $\times$ \\
    $\tau_y \otimes s_z \otimes \sigma_x$ & $\tau_y \otimes \sigma_y$ & 0 & 0 & $\times$ & $\times$\\
    $\tau_y \otimes s_z \otimes \sigma_z$ & $\tau_y \otimes \sigma_x$ & 0 & 0 & $\checkmark$ & $\times$ \\
    $\tau_x \otimes s_z \otimes \sigma_y$ & 0 & $-\tau_y \otimes \sigma_y$ & $\tau_x \otimes \sigma_0$ & $\times$ & $\times$ \\
    \hline
\end{tabular}
\caption{The effective projection of the $C_2$ allowed coupling terms on (100), (010), and (001) surfaces. ``$\checkmark$'' means the surface is gapped, while ``$\times$'' means the surface is gapless.}
\label{c2 table}
\end{table}

\subsection{$C_4$}

Now we consider the case with $C_4$, where the band insulator includes two single TIs with $l_z= \frac{1}{2}$ and $\frac{3}{2}$. The unperturbed zero-mode eigenfunctions are the same as those given in Eqs.~\ref{eq:001-surf-states} and \ref{eq:100-surf-states}. Here we consider the quadratic-in-$k$ couplings between two single TIs:
\begin{align}
    H_c({\bf k}) = (k_y^2-k_x^2)\tau_{x,y} \otimes s_{x,y} \otimes \sigma_{0,x,y,z},
\end{align} 
which respect $C_4=\frac{\sqrt{2}}{2} (\tau_z \otimes s_0 \otimes \sigma_0 - i\tau_0 \otimes s_z \otimes \sigma_0)$. Further the TRS ($T=i\tau_0 \otimes s_y \otimes \sigma_0 K$) allows only 8 coupling forms:
\begin{align}
    \tau_{y} \otimes s_{y} \otimes \sigma_{0,x,z}, \quad \tau_{x} \otimes s_{y} \otimes \sigma_{y}, \quad \tau_{y} \otimes s_{x} \otimes \sigma_{0,x,z}, \quad \tau_{x} \otimes s_{x} \otimes \sigma_{y},
\end{align}
and the projections of these terms on (100),(010) and (001) surfaces are shown in Table.~\ref{c4 table}. First, we note the four non-zero couplings on (001) vanish along the high-symmetry path $k_x = \pm k_y$, leaving two pairs of Dirac cones centered at finite $k$. So the (001) surface always remains gapless.

When projecting $H_c$ onto the side surfaces, the $k^2$ terms will act as spatial derivatives due to open boundary conditions. For instance, for $(100)$ surface,
\begin{align}
	k_x^2 \rightarrow F_x=\int_{-\infty}^0 dx \psi^*(x)(-i\partial_x)^2\psi(x) = \alpha^2+\beta^2 \quad , \quad \psi(x) = f(x)\xi = \mathcal{N} sin(\alpha x)e^{\beta x} \xi.
\end{align}
where $\alpha = \frac{\sqrt{-4m_0m_2-B}}{2m_2}, \beta = \frac{B}{2m_2}$ and the normalization factor $\mathcal{N}=\sqrt{4\beta(\alpha^2+\beta^2)/\alpha^2}$. The wavefunction $\psi(x)$ is obtained following steps similar to those in Eq.~\ref{wavefunction on surface}, under the condition $f(x=0)=f(x=-\infty)=0$. 

Since $H_\pm$ are energetically shifted by an energy scale $\delta$, the intersection of their Dirac surface states is a closed ring in $k$ space with a radius of $\delta/B$, around which the coupling $H_c$ will open up a gap. Taken together, the projected surface gap, if non-zero, is generally proportional to $F_x - \eta$, where we have defined $\eta=(\delta/B)^2$. In our case, $B=m_2=-m_0=1 > \delta$. Thus, $F_x=1 \gg \eta$ ensures $F_x - \eta \neq 0$. For the spin component, only $\tau_y \otimes \sigma_x$ anticommutes with both $\tau_z\otimes \sigma_x$ and $\tau_0 \otimes \sigma_y$, allowing a quadratic coupling that gaps the (100) surface. The same analysis holds for the (010) surface, we see that $(k_y^2-k_x^2)\tau_y \otimes s_y \otimes \sigma_x$ can gap the whole side surfaces and leaves (001) gapless. Under the original basis $( |l_z\rangle , |-l_z \rangle, |l_z \rangle , |-l_z\rangle )^T$ this term reads $(k_y^2-k_x^2)\tau_y \otimes s_x \otimes \sigma_y$.

Similar to the $C_2$ case, the $C_4$-respecting double TI can trap four pairs of gapless helical hinge modes related by $C_4$. The above discussion can be naturally extended to systems with $C_6$.

\begin{table}[t]
\centering
\begin{tabular}[c]{|c|ccc|c|c|}
    \hline
    Coupling terms & (100) & (010) & (001) &\text{Gap Presence on (100)} & \text{Gap Presence on (010)} \\
    \hline
    $(k_y^2-k_x^2)\tau_y \otimes s_y \otimes \sigma_x$ & $({\eta-F_x})\tau_y \otimes \sigma_x$ & $(\eta-F_y)\tau_x \otimes \sigma_0$& $ (k_x^2-k_y^2)\tau_y \otimes \sigma_x$ & $\checkmark$ & $\checkmark$ \\
    $(k_y^2-k_x^2)\tau_y \otimes s_y \otimes \sigma_z$ & $(F_x-\eta) \tau_y \otimes \sigma_y$ & $(F_y-\eta) \tau_y \otimes \sigma_z$ & $(k_y^2-k_x^2) \tau_y \otimes \sigma_y$ & $\times$ & $\times$ \\
    $(k_y^2-k_x^2)\tau_y \otimes s_y \otimes \sigma_0$ & 0 & 0 & 0 & $\times$ & $\times$ \\
    $(k_y^2-k_x^2)\tau_x \otimes s_y \otimes \sigma_y$ & 0 & 0 & 0 & $\times$ & $\times$ \\
    $(k_y^2-k_x^2)\tau_y \otimes s_x \otimes \sigma_x$ & 0 & 0 & $(k_y^2-k_x^2) \tau_y \otimes \sigma_y$ & $\times$ & $\times$ \\
    $(k_y^2-k_x^2)\tau_y \otimes s_x \otimes \sigma_z$ & 0 & 0 & $(k_y^2-k_x^2) \tau_y \otimes \sigma_0$ & $\times$ & $\times$\\
    $(k_y^2-k_x^2)\tau_y \otimes s_x \otimes \sigma_0$ & $(\eta-F_x) \tau_y \otimes \sigma_z$ & $(F_y-\eta) \tau_y \otimes \sigma_y$ & 0 & $\times$ & $\times$ \\
    $(k_y^2-k_x^2)\tau_x \otimes s_x \otimes \sigma_y$ & $(\eta-F_x) \tau_x \otimes \sigma_0$ & $(F_y-\eta) \tau_y \otimes \sigma_x$ & 0 & $\times$ & $\times$ \\
    \hline 
\end{tabular}
\caption{$C_4$-invariant perturbations and their gapping capabilities on (100), (010), and (001) surfaces. The ``$\checkmark$'' means the surface is gapped, while ``$\times$'' means the surface is gapless. }
\label{c4 table}
\end{table}

\section{Spatial localization of hinge modes}
In this section, we discuss the localization properties of hinge states in our TCI model. These hinge states are expected to decay as $e^{-\frac{r}{\lambda}}$, where $\lambda$ is the localization length. 

\begin{figure}[t]
     \includegraphics[scale=0.4]{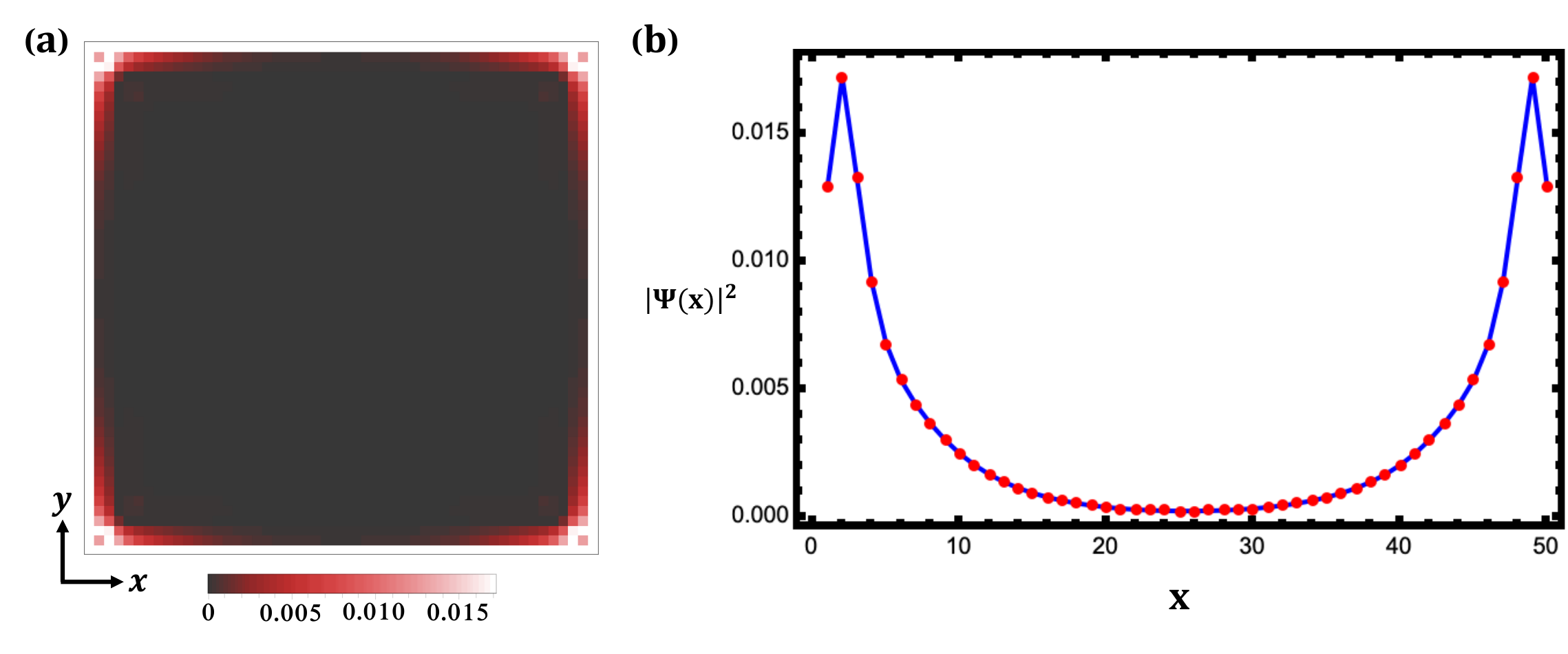}
     \caption{(a) The local density distribution of hinge states, $|\Psi(x,y)|^2$. (b) The local density distribution along $y=1$, $|\Psi(x)|^2$.}
     \label{Fig.A4}
 \end{figure}

For the $C_4$-invariant TCI model in the main text, we place it on a $50\times50$ square lattice with the parameter set $A=0.5$ and $ B=\delta=m_1=m_2=-m_0=1$. For coupling strength $g=0.1$, the local density of states of the four hinge states is shown in Fig.~\ref{Fig.A4} (a). Fig.~\ref{Fig.A4} (b) shows $|\Psi(x)|^2$, the density distribution along $y=1$, as a function of the lattice position $x$. By fitting the logarithm of $|\Psi(x)|^2$, we find $\lambda \approx 5$, in units of lattice constants. Note that $\lambda$ is inversely proportional to the surface gap or the coupling strength $g$. In realistic systems, we expect $g\sim 0.01$ (e.g., in units of eV) and $\lambda \sim 50$ lattice constants. This should correspond to a hinge-state localization length of a few dozen nanometers.

\section{Superconducting gap and flux insertion}

In order to achieve a fully gapped TSC state under a non-zero magnetic flux, the s-wave pairing order should carry a vortex phase winding $e^{-i\theta}$~\cite{de2019conditions}. As a result, the angular momentum eigenvalue $j_z$ associated with the flux mode will be shifted to $0$ or $n/2$, depending on the helicity of the TI Hamiltonian. In this section, we will construct an effective theory for different forms of superconducting pairing based on the flux modes from a single TI nanowire, thereby demonstrating the necessity of phase introduction and the way angular momentum is shifted at $\pi$-flux.

We first discuss the TI nanowire with positive helicity, $H_+({\bf k})$. The eigenfunctions of the boundary state have been derived in Appendix B:
\begin{align}
    \psi_{1+}^e = \frac{e^{im_1\theta}}{\sqrt{2}} \begin{bmatrix}
		1 \\ 0 \\0 \\ie^{i\theta}
	\end{bmatrix} \quad , \quad \psi_{2+}^e = \frac{e^{-i(m_2+1)\theta}}{\sqrt{2}} \begin{bmatrix}
		0  \\ -e^{i\theta} \\ -i \\ 0
	\end{bmatrix}
\end{align}
and the energy of the two states are $E_{1,2}^e=\frac{B}{R}(m_{1,2}+\frac{1}{2})$. Since TRS is respected under zero-flux, the Kramers' theorem asks a band distribution with double degeneracy. The $s$-wave pairing order is $-\Delta_0 \tau_y \otimes s_0 \otimes \sigma_y$ and the corresponding hole states $\psi_i^h = (\psi_i^e)^*$ have opposite energy $E_{1,2}^h=-\frac{B}{R}(m_{1,2}+\frac{1}{2})$. We now construct a low-energy Nambu basis using two degenerate electron states and their corresponding hole states:
\begin{align}
    \Psi_1 \equiv (\psi_{1+}^e,0)^T  \quad , \quad   \Psi_2 \equiv 
	    (\psi_{2+}^e,0)^T  \quad , \quad \Psi_3 \equiv (0,\psi_{1+}^{e*})^T  \quad , \quad   \Psi_4 \equiv 
	    (0,\psi_{2+}^{e*})^T .
\end{align}
The projection of the superconducting pairing onto the above subspace is:
\begin{align}
     \Delta_{ij}^{l=0}(\theta)= \langle \Psi_i | -\Delta_0 \tau_y \otimes s_0 \otimes \sigma_y | \Psi_j \rangle &= \begin{bmatrix}
       0 & 0 & 0 & -\Delta_0 \\
       0 & 0 & \Delta_0 & 0 \\
       0 & \Delta_0 & 0 & 0 \\
       -\Delta_0 & 0 & 0 & 0
    \end{bmatrix} = \Delta_0 s_y\otimes \sigma_y.
\end{align}
The effective pairing is a constant matrix and thus it can open a uniform superconducting gap. Since $\psi_1^e$ and $\psi_2^h$ carry the same angular momentum, $j_z = m+\frac{1}{2}$, it is natural for the effective pairing between them to be nonvanishing, so do $\psi_2^e$ and $\psi_1^h$ with $j_z=-m-\frac{1}{2}$. This essentially reflects the constraint imposed by angular momentum conservation.

Next we introduce a magnetic flux, the eigenstate remains unchanged but the energy level shifts due to the breaking of TR symmetry (according to Appendix C):
\begin{align}
    \Psi_{1,l} \equiv \Psi_1 \quad &, \quad E_{1,l}^e = \frac{B}{R}(m_1+\frac{1}{2}+l) \quad , \quad j_1^e = m_1 + \frac{1}{2}. \nonumber \\
    \Psi_{2,l}\equiv \Psi_2 \quad &, \quad E_{2,l}^e = \frac{B}{R}(m_2+\frac{1}{2}-l) \quad , \quad j_2^e = -(m_2 + \frac{1}{2}). \nonumber \\
    \Psi_{3,l} \equiv \Psi_3 \quad &, \quad E_{1,l}^h =-\frac{B}{R}(m_1+\frac{1}{2}+l) \quad , \quad j_1^h = -(m_1+\frac{1}{2}). \nonumber \\
    \Psi_{4,l} \equiv \Psi_4 \quad &, \quad E_{2,l}^h = -\frac{B}{R}(m_2+\frac{1}{2}-l) \quad , \quad j_2^h = m_2+\frac{1}{2}.
\end{align}
In the presence of a nonzero flux (including $\pi$-flux), the previously degenerate energy bands become misaligned, allowing a nontrivial superconducting gap to open at the Fermi surface only through the coupling between electrons and their corresponding hole states. However, an analysis of the angular momentum reveals an immediate inconsistency: for any integer $m_1$ and $m_2$, $j_1^e \neq j_1^h$ and $j_2^e \neq j_2^h$. This rotational symmetry-hindered pairing is also reflected in the effective projection: 
\begin{align}
    \Delta_{ij}^{l}(\theta) &= \langle \Psi_{i,l} | -\Delta_0 \tau_y \otimes s_0 \otimes \sigma_y | \Psi_{j,l} \rangle \nonumber \\
    &= \begin{bmatrix}
       0 & 0 & 0 & -\Delta_0 e^{i(m_2-m_1)\theta} \\
       0 & 0 & \Delta_0  e^{i(m_2-m_1)\theta} & 0 \\
       0 & \Delta_0  e^{i(m_1-m_2)\theta} & 0 & 0 \\
       -\Delta_0  e^{i(m_1-m_2)\theta} & 0 & 0 & 0
    \end{bmatrix}.
\end{align}
Formally, although there are still some finite matrix elements of $\Delta_{ij}^{l}=\int_{0}^{2\pi}d\theta \Delta_{ij}^{l}(\theta)$ allowed by symmetry, the resulting projected pairing can be non-zero on the Fermi surface. Specifically, $j_1^e=j_2^h$ but $E_1^e \neq -E_2^h$ in the case of $m_1=m_2$, so the non-zero component does not guarantee the existence of a global superconducting gap. For the special $\pi$-flux case, the effective pairing between two degenerate states is:
\begin{align}
   \Delta_{ij}^{l=\frac{1}{2}} =\int_{0}^{2\pi}d\theta  \Delta_{ij}^{l=\frac{1}{2}}(\theta) = \int_{0}^{2\pi}d\theta \begin{bmatrix}
       0 & 0 & 0 & -\Delta_0 e^{-i\theta} \\
       0 & 0 & \Delta_0  e^{-i\theta} & 0 \\
       0 & \Delta_0  e^{i\theta} & 0 & 0 \\
       -\Delta_0  e^{i\theta} & 0 & 0 & 0
    \end{bmatrix} = 0,
\end{align}
which indicates the absence of a pairing gap on the entire Fermi surface.

To resolve the angular momentum mismatch, it is necessary to introduce an additional superconducting vortex phase. The $s$-wave pairing is now modified to the following form under Nambu basis $( \Psi_e , \Psi_h)^T$:
\begin{align}
    \begin{bmatrix}
        0 & \Delta_0 e^{-in_\nu\theta} \\
        \Delta_0^{\dagger}e^{in_\nu\theta} & 0 
    \end{bmatrix} = U^{-1}\begin{bmatrix}
        0 & \Delta_0  \\
        \Delta_0^{\dagger} & 0 
    \end{bmatrix} U  \quad , \quad U=\begin{bmatrix}
        e^{in_\nu \frac{\theta}{2}} & 0 \\
        0 & e^{-in_\nu \frac{\theta}{2}}
    \end{bmatrix}.
\end{align}
$U$ can also be viewed as the active transformation of the basis vectors. This results in a $n_\nu/2$ shifting in the angular momentum, but with opposite signs for electrons and holes. The eigenstates of the electron and hole after the gauge transformation are:
\begin{align}
    \Psi_{1,l}^{n_{\nu}} \equiv U\Psi_{1,l} =e^{i\frac{n_\nu}{2}\theta} \Psi_{1,l} \quad , \quad E_{1,l}^e = \frac{B}{R}(m_1+\frac{1}{2}+l) \quad &, \quad j_{1,n_\nu}^e = m_1 + \frac{1}{2}+\frac{n_v}{2}. \nonumber \\
    \Psi_{2,l}^{n_{\nu}}\equiv U\Psi_{2,l}  = e^{i\frac{n_v}{2}\theta} \Psi_{2,l} \quad, \quad E_{2,l}^e = \frac{B}{R}(m_2+\frac{1}{2}-l) \quad &, \quad j_{2,n_\nu}^e = -(m_2 + \frac{1}{2}-\frac{n_v}{2}). \nonumber \\
    \Psi_{3,l}^{n_{\nu}} \equiv U\Psi_{3,l} = e^{-i\frac{n_\nu}{2}\theta}\Psi_{3,l} \quad , \quad E_{1,l}^h =-\frac{B}{R}(m_1+\frac{1}{2}+l) \quad &, \quad j_{1,n_\nu}^h = -(m_1+\frac{1}{2}+\frac{n_\nu}{2}). \nonumber \\
    \Psi_{4,l}^{n_\nu} \equiv U\Psi_{4,l} = e^{-i\frac{n_\nu}{2}\theta}\Psi_{4,l} \quad , \quad E_{2,l}^h = -\frac{B}{R}(m_2+\frac{1}{2}-l)\quad &, \quad j_{2,n_\nu}^h = m_2+\frac{1}{2}-\frac{n_\nu}{2}.
\end{align}
With $l=1/2,n_\nu=1$, it is easy to see that electron and hole states at the same energy must carry the same angular momentum as well. This enables a finite superconducting gap in the system. 

We check this by projecting the BdG Hamiltonian onto the subspace of the helical mode with $m_1=-1$ and $m_2=0$:
\begin{align}
    (H_{+})_{\text{eff}} &= \langle \Psi_{i,l=\frac{1}{2}} |  Ak_z \tau_0 \otimes s_x \otimes \sigma_z -\mu\tau_z \otimes s_0 \otimes \sigma_0 -\Delta_0 \tau_y \otimes s_0 \otimes \sigma_y | \Psi_{j,l=\frac{1}{2}} \rangle \nonumber \\
    & = \begin{bmatrix}
       -\mu & iAk_z & 0 & -\Delta_0 \\
       -iAk_z & -\mu & \Delta_0 & 0 \\
       0 & \Delta_0 & \mu & -iAk_z \\
       -\Delta_0 & 0 & iAk_z & \mu
    \end{bmatrix} = \Delta_0 s_y \otimes \sigma_y - Ak_z s_z \otimes \sigma_2 - \mu s_z \otimes \sigma_0 ,
\end{align}
where $\mu$ is the chemical potential. The energy dispersion of this effective Hamiltonian is $E=\pm\sqrt{(Ak_z\pm\mu)^2+\Delta
_0^2}$, indicating a general superconducting gap $\Delta_0$ for any value of $\mu$. 

\begin{figure}[t]
    \includegraphics[scale=0.5]{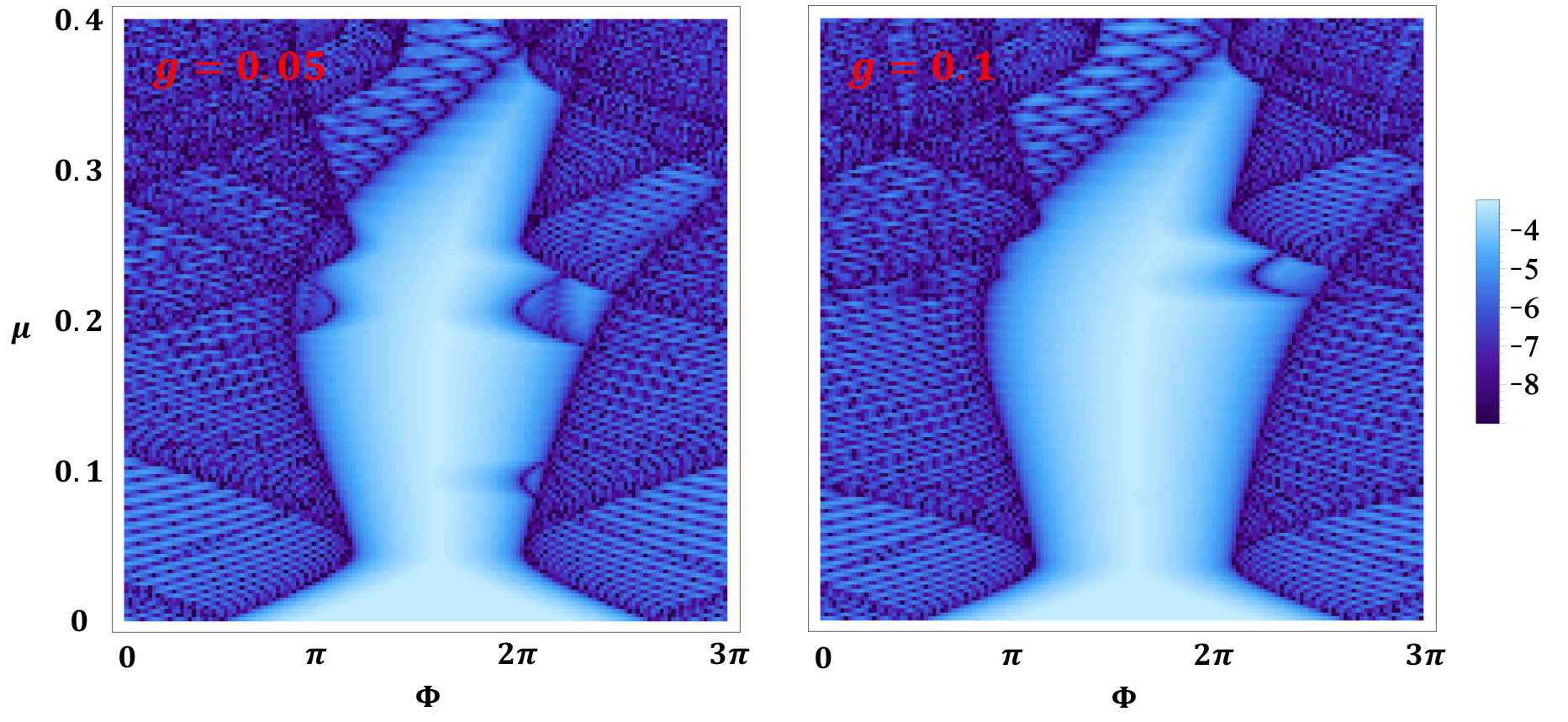}
    \caption{The topological phase diagram of $H_{BdG}$ as a function of $\mu$ and $\Phi$ with different inter-TI coupling strength, g=0.05 (left) and 0.1 (right). The color denotes the logarithm of the BdG energy gap.}
    \label{Fig.A5}
\end{figure}

The same analysis method applies to the TI nanowire with negative helicity, $H_{-}{({\bf k})}$. The wavefunctions and eigenvalues of the helical mode with both flux and superconducting vortex are given here:
\begin{align}
    \Phi_{1,l}^{n_{\nu}} \equiv e^{i\frac{n_\nu}{2}\theta} (\psi_{1-}^e,0)^T \quad , \quad E_{3,l}^e = -\frac{B}{R}(m_3-\frac{1}{2}+l)\quad &, \quad j_{3,n_\nu}^e = (m_3 - \frac{1}{2}+\frac{n_v}{2})+\frac{n}{2}. \nonumber \\
    \Phi_{2,l}^{n_{\nu}} \equiv  e^{i\frac{n_\nu}{2}\theta} (\psi_{2-}^e,0)^T \quad, \quad E_{4,l}^e = -\frac{B}{R}(m_4-\frac{1}{2}-l)\quad &, \quad j_{4,n_\nu}^e = -(m_4 - \frac{1}{2}-\frac{n_v}{2})+\frac{n}{2}. \nonumber \\
    \Phi_{3,l}^{n_{\nu}} \equiv e^{-i\frac{n_\nu}{2}\theta} (0,\psi_{1-}^{e*})^T \quad, \quad E_{3,l}^h =\frac{B}{R}(m_3-\frac{1}{2}+l) \quad&, \quad j_{3,n_\nu}^h = -(m_3-\frac{1}{2}+\frac{n_\nu}{2})-\frac{n}{2}. \nonumber \\
    \Phi_{4,l}^{n_{\nu}} \equiv e^{-i\frac{n_\nu}{2}\theta} (0,\psi_{2-}^{e*})^T \quad, \quad E_{4,l}^h = \frac{B}{R}(m_4-\frac{1}{2}-l) \quad &, \quad j_{4,n_\nu}^h = (m_4-\frac{1}{2}-\frac{n_\nu}{2})-\frac{n}{2}.
\end{align}
where the eigenfunctions $\psi_{3,4}^e$ are derived in Appendix B and C. Consistent with the results of the symmetry argument, the helical mode with $m_3=0$ and $m_4=1$ under the $\pi$-flux ($l=1/2, n_\nu=1$) carries $j_3^e=j_4^e=n/2$. The angular momentum of the hole state is $j_3^h=j_4^h=-n/2$, which is equivalent to $n/2$, so the conservation condition of angular momentum is still satisfied. The effective BdG Hamiltonian in the case of $H_{-}({\bf k})$ is : $Ak_z s_z \otimes \sigma_2 - \mu s_z \otimes \sigma_0 + \Delta_0 s_y \otimes \sigma_y$ and the energy dispersion is the same: $E=\pm\sqrt{(Ak_z\pm\mu)^2+\Delta_0^2}$.

Finally, we numerically map out the topologically superconducting phase diagram and check its dependence on the ``inter-TI" coupling strength $g$. Similar to Fig.~3 (a) in the main text, the phase diagram here is obtained by evaluating the logarithm of the global gap, log($\Delta_{gap}$), of $H_{BdG}$. To examine the effect of varying coupling strengths, Fig.~\ref{Fig.A5} shows phase diagrams computed for different $g$ values. The results indicate that, qualitatively, increasing $g$ enlarges the TSC phase region and blurs the phase boundary between two single TIs. This indicates that an increase of $g$ helps stabilize the TSC phase.

\end{document}